\documentclass[12pt,twoside]{amsart}
\usepackage{amssymb}
\usepackage[english]{babel}
\usepackage[utf8]{inputenc}
\usepackage{natbib}
\usepackage{comment}
\usepackage{tikz}
\usetikzlibrary{calc}
\usepackage{graphicx}
\usepackage{a4wide}
\def\RR{\mathbb{R}}    %
\def\EE{\mathbb{E}}    %
\def\R{\mathcal{R}}   %

\def\proof{\noindent{\em Proof.}~}
\def\eproof{\mbox{\ }\hfill$\square$}
\newtheorem{theorem}{Theorem}
\newtheorem{lemma}{Lemma}
\newtheorem{proposition}{Proposition}
\newtheorem{corollary}{Corollary}
\newtheorem{assumption}{Assumption}
\newtheorem{definition}{Definition}
\newtheorem{example}[theorem]{\textsc{Example}}
\newtheorem{remark}[theorem]{Remark}
\date{}

\title{Distributionally Robust Rational Inattention}

\author{Emerson Melo}

\date{\today}

\calclayout
\calclayout
\thanks{This paper is a substantially revised version of \emph{On the Distributional Robustness of Finite Rational Inattention Models} (arXiv:2208.03370). Department of Economics,
	Indiana University, Bloomington, IN 47408, USA. Email: {\tt emelo@iu.edu}. I am very grateful to Bob Becker, Austin Knies,  and Jorge Lorca  for their  valuable comments and suggestions that have greatly improved the paper.}
\begin{document}

\begin{abstract}
	We study a rational inattention model in which a decision maker chooses a stochastic choice rule while facing ambiguity about the prior distribution over states. Ambiguity is modeled through a distributionally robust formulation in which the prior belongs to an uncertainty set defined by a general $\phi$-divergence. Our main result establishes an exact Lagrangian dual representation of the resulting max--min problem as a finite-dimensional concave optimization problem. This characterization yields necessary and sufficient conditions for the emergence of robust consideration sets and a tractable one-dimensional dual representation for the Cressie--Read family of divergences. We further connect the model to robust risk measures: Average Value-at-Risk arises from a particular non-smooth divergence, while Kullback--Leibler ambiguity is linked through Lagrangian duality to the entropic-risk multiplier representation. We then develop two extensions. The first allows the attention-cost technology and prior ambiguity to be governed by distinct $\phi$-divergences, while the second introduces a deeper layer of ambiguity over the decision maker's baseline prior. Both extensions preserve the finite-dimensional structure of the model and encompass standard Kullback--Leibler robust rational inattention specifications as special or corresponding multiplier cases.
\end{abstract}

	\maketitle
	
	\pagestyle{myheadings} \thispagestyle{plain} \markboth{ }{ }
	
	\vspace{3ex}
	
	\thispagestyle{empty}
	
	\thispagestyle{empty}
	
	\newcommand{\spacing}[1]{\renewcommand{\baselinestretch}{#1}\large\normalsize}
	\textwidth      5.95in \textheight 600pt
	\spacing{1.1}%
	\emergencystretch=3em
	
	\section{Introduction}\label{s1}
	Rational Inattention (RI) studies decision makers who choose how much information to acquire before acting when learning is costly (\cite{Sims2003,Sims2010}). Rather than taking an information structure as given, the decision maker optimally selects one subject to an information-processing constraint. Information is therefore allocated selectively: attention is directed toward payoff-relevant uncertainty while less valuable information is ignored.
	\smallskip

	A central result of \cite{Matejka2015} shows that, under Shannon mutual information costs, the optimal choice rule takes the form of a generalized multinomial logit (MNL), with choice probabilities determined jointly by payoffs and the decision maker's prior beliefs over states.\footnote{A prior is a probability measure over the set of states of the world.} Like most of the subsequent RI literature, however, this framework assumes that the prior is known. In the terminology of \cite{Hansen_Sargent_2001,Hansen_Sargent2008,Marinaccietal2006}, the decision maker faces neither prior ambiguity nor model misspecification.\footnote{This assumption is also maintained in empirical applications of RI; see, for example, \cite{DASGUPTA2018150}, \cite{BrownJeon2021}, \cite{Porcher2020}, \cite{Joo2019}, and \cite{Lugovski_et_al_2021}.} Thus, although RI relaxes what the decision maker knows about payoffs, it continues to assume precise beliefs about the distribution of states.
	\smallskip

	This paper studies rational inattention when the decision maker does not fully trust her prior beliefs over states. Rather than assuming a known prior distribution, we allow the prior to vary within an uncertainty set defined by a general $\phi$-divergence (\cite{LieseVajda1987}) around a benchmark prior $\mu$. The decision maker therefore chooses an information structure while guarding against the worst-case prior in this ambiguity set. Exploiting the characterization of optimal rational inattention as a stochastic optimization problem over unconditional choice probabilities (\cite[Cor.~1]{Matejka2015}), we formulate the resulting problem as a distributionally robust optimization problem, which we call Distributionally Robust Rational Inattention (DRO-RI). Our main result shows that the DRO-RI problem admits an equivalent finite-dimensional concave formulation, providing a tractable characterization of optimal behavior under general $\phi$-divergence ambiguity.
		\smallskip

	Combining costly information acquisition with prior ambiguity is natural whenever decision makers must learn about environments whose underlying probability distributions are themselves uncertain. In such settings, the decision maker faces two distinct frictions: acquiring information is costly, and the probabilistic model used to evaluate that information may itself be misspecified. Existing rational inattention models account for the former but typically abstract from the latter. 
	
	These two frictions arise jointly in a variety of economic settings. In portfolio choice and financial intermediation, investors allocate costly attention across assets while remaining uncertain about the return-generating process (\cite{HuangLiu2007,Hansen_Sargent_2001,Hansen_Sargent2008,Kasa2006,HanKasaLuo2024}). Likewise, insurance consumers devote limited attention to contract details (\cite{BrownJeon2021}) while facing ambiguity about the underlying loss distribution (\cite{AlaryGollierTreich2013}), particularly for novel or rapidly evolving risks such as cyber insurance or climate-related losses. Recent work by \cite{Fabbri2024} combines rational inattention with smooth ambiguity preferences. In that model the information cost is evaluated at a fixed predictive prior, independent of ambiguity attitudes, which produces state-dependent reward multipliers in the optimal choice rule and violates invariance under compression. In our model, as in \cite{Hansen_et_al_2022}, the cost is evaluated at the adversarial prior, so the Matejka--McKay logit structure is preserved under the worst-case prior. More generally, there is currently no tractable characterization of rational inattention under general $\phi$-divergence ambiguity --- the gap this paper fills.

	\smallskip

Our first contribution is an exact characterization of the DRO-RI problem. We show that, for an arbitrary $\phi$-divergence, the original max--min optimization problem is equivalent to a finite-dimensional concave program (Theorem~\ref{Thm_DRO-RI-phi}). This dual characterization preserves the tractability of the classical RI problem while jointly determining the optimal choice probabilities and the worst-case prior. As a consequence, the optimal choice rule retains a generalized MNL structure, extending the characterization of \cite{Matejka2015} to environments with prior ambiguity.
	\smallskip

	Our second contribution is to provide an economic interpretation of DRO-RI through the theory of convex and coherent risk measures (\cite{Shapiro_et_al_2013}). We show that the choice of $\phi$-divergence determines how the decision maker evaluates ambiguity over priors. In particular, the multiplier (penalty) form of Kullback--Leibler ambiguity induces entropic risk, to which the constrained Kullback--Leibler ball is linked through Lagrangian duality; other divergences recover Average Value-at-Risk and, for the Cressie--Read family, a characterization in terms of tail performance under the worst-case prior. To our knowledge, this connection between rational inattention and risk-measure theory is new.
		\smallskip

	Our third contribution is to extend the framework to allow separate sources of robustness in information acquisition and prior beliefs. Specifically, we allow the attention technology to be robustified through a $\phi_1$-divergence while the ambiguity set over priors is governed by a (possibly different) $\phi_2$-divergence. Building on the composite representation of robust risk measures developed by \cite{Jinetal2024}, we show that this composite DRO-RI model also admits an equivalent finite-dimensional concave formulation (\S\ref{s34}). The framework nests both the penalized (multiplier) counterpart of our single-divergence model and the model of \cite{Hansen_et_al_2022} as special cases, corresponding respectively to $\phi_1=\phi_{kl}$ and $\phi_1=\phi_2=\phi_{kl}$. This allows the information-processing technology and the decision maker's beliefs to exhibit distinct forms of model uncertainty, substantially broadening the class of robust rational inattention models that can be analyzed.
	
	\subsection{Related literature}\label{s11}
	
	Our paper contributes to the literature on rational inattention and discrete choice initiated by \cite{Matejka2015}. Subsequent work has generalized the information-cost function (\cite{Fosgerauetal2020}, \cite{Caplin_et_al2022}, \cite{MorrisYang_2021}, \cite{Woodford_Hebert_2021}), characterized endogenous consideration sets (\cite{Caplinetal2019}), and developed geometric foundations for RI (\cite{Armenteretal2021}).\footnote{For a comprehensive survey of rational inattention, see \cite{Matejkaetal2022}.} Throughout this literature, however, the decision maker's prior is taken as known. Our paper relaxes this assumption by incorporating distributional ambiguity over the prior while preserving the tractability of finite rational inattention models.
	\smallskip

	The paper most closely related to ours is \cite{Hansen_et_al_2022}. Like us, they study robust RI in finite environments, but they formulate robustness through a Kullback--Leibler multiplier preference following the robust-control framework of \cite{Hansen_Sargent_2001,Hansen_Sargent2008}. By contrast, our baseline model adopts a constraint formulation based on general $\phi$-divergence balls. In the Kullback--Leibler case, the associated optimal Lagrange multiplier provides the correspondence with the multiplier formulation studied by \cite{Hansen_et_al_2022}.
	\smallskip
	
	Relative to \cite{Hansen_et_al_2022}, our framework differs along four dimensions: the generality of the divergence class; tractable characterizations for the Cressie--Read family, including hard attention thresholds (exact zero conditional choice probabilities) generated by Cressie--Read attention costs, which are unattainable under Shannon costs, and worst-case priors without full support generated by Cressie--Read or AV$@$R ambiguity, which are unattainable under Kullback--Leibler ambiguity; the explicit risk-measure connection identifying their model as our entropic-risk special case; and the composite extension of \S\ref{s34}, which allows separate forms of robustness in the attention technology and the prior. For their part, \cite{Hansen_et_al_2022} develop a numerical algorithm applicable to arbitrary utility functions, whereas our emphasis is on analytical characterization under broad classes of divergence measures. The two approaches are therefore complementary.
	
		\smallskip

	More broadly, our paper connects the literatures on ambiguity-sensitive decision making (\cite{GILBOA1989141}, \cite{Marinaccietal2006}, \cite{Strzalecki2011}; see \cite{Marinacci_2015} for a survey) and distributionally robust optimization (\cite{Shapiro,Shapiro_et_al_2021.Ch7,Kuhn_et_al_2019}) with the theory of rational inattention. Closest to our nested extension in \S\ref{s36} is \cite{CerreiaVioglioetal2026}, who axiomatize a criterion that combines max--min ambiguity over a set of structured models with a \cite{Hansen_Sargent_2001}-type concern that all of these models are misspecified; their decision maker, however, chooses among acts under full information, whereas ours also chooses what to learn. While these literatures have developed largely independently, our framework combines them within a unified model of costly information acquisition under prior ambiguity. Our composite extension further builds on the two-divergence representation of robust risk measures developed by \cite{Jinetal2024} and the optimized certainty equivalent of \cite{BenTalTeboulle2007}. To the best of our knowledge, this is the first application of these ideas to rational inattention.
		\smallskip

	The remainder of the paper is organized as follows. Section~\ref{s2} presents the rational inattention framework and introduces prior ambiguity. Section~\ref{s3} formulates the Distributionally Robust Rational Inattention (DRO-RI) problem, develops its finite-dimensional concave characterization, and connects it to risk measures. Section~\ref{s45} develops the composite (two-divergence) and nested extensions. Section~\ref{s44} characterizes robust consideration sets and illustrates them under Kullback--Leibler, AV$@$R, and Cressie--Read specifications. Section~\ref{s6} concludes. All proofs and supplementary technical results are collected in Appendix~\ref{Proofs}.

	\section{The RI model }\label{s2}
	We consider the finite rational inattention model of \cite{Matejka2015}. A decision maker (DM) chooses one alternative from a finite set $A=\{1,\ldots,n\}$ under a finite state space $\Theta\subseteq\mathbb{R}^d$ with $m\triangleq|\Theta|$ states. Utility depends on the realized state $\theta\in\Theta$, which is unobserved, so the payoff vector $u(\theta)=(u(a,\theta))_{a\in A}$ is unknown when the decision is made.

	 Throughout this section we adopt the classical rational inattention assumption that the decision maker's prior $\mu\in\Delta(\Theta)$ correctly describes the distribution of states.
	\begin{assumption}[No model uncertainty]\label{No_Model_uncertainty} The DM possesses correct prior knowledge about the available options, given by a probability measure $\mu\in \Delta(\Theta)$, where $\mu(\theta)\triangleq \mathbb{P}(\tilde\theta=\theta)>0$ for all $\theta \in \Theta$, and $\tilde\theta$ denotes the random state.
	\end{assumption}

	Let $p_a(\theta)$ be the conditional probability of choosing $a$ given $\theta$, $p(\theta)=(p_a(\theta))_{a\in A}$, $P=\{p(\theta)\}_{\theta\in\Theta}\in\mathcal P$. Given $P$ and $\mu$, the \emph{unconditional} choice probability vector is $x=(x_a)_{a\in A}$, $x_a=\sum_{\theta\in\Theta}p_a(\theta)\mu(\theta)$. The DM's problem has two stages: she first chooses an information strategy that refines $\mu$, at a cost increasing in informativeness, and then makes a standard decision under the beliefs it generates.
	\smallskip

	A fundamental result in \cite{Matejka2015} shows this problem can be written as \emph{directly} choosing $P$, once information costs are modeled by the Shannon mutual information:
	\begin{eqnarray}\label{Mutual_information_cost}
		\xi C(P, \mu) &\triangleq&\xi\left[\sum_{a\in A}\sum_{\theta\in\Theta}p_{a}(\theta)\left(\log p_{a}(\theta)-\log x_{a}\right)\mu(\theta)\right], \notag\label{MutualEntropy}
	\end{eqnarray}
	where $\xi>0$ is the information-cost parameter.
		\smallskip

	Following \cite{Matejka2015} and \cite{Fosgerauetal2020}, the DM solves
	\begin{equation}
		\max_{P\in \mathcal{P}}\left\{ \sum_{\theta\in \Theta}	\left[\sum_{a\in A}
		p_{a}(\theta)u(a,\theta)\right]\mu(\theta) -\xi C(P, \mu) \right\} ,
		\label{Program1}
	\end{equation}
	subject to
	\begin{equation}
		p_{a}(\theta)\geq 0\ \text{for all $a\in A$}\quad \text{and} \quad\sum_{a\in A}p_{a}(\theta)=1\quad \forall \theta\in \Theta.  \label{constrain2}
	\end{equation}

	As \cite{Matejka2015} show, a solution is given by the generalized MNL formula
	\begin{equation}  \label{RIlogitprobs}
		p_{a}(\theta)=\frac{x_a e^{u(a,\theta)/\xi}}{\sum_{b\in A}x_be^{u(b,\theta)/\xi
		}}\quad \text{for all $a\in A$,}
	\end{equation}%
	in which the prior's entire influence on choice is captured by the shift term $\log x_a$, independent of realized payoffs. Recovering $P$ therefore reduces to solving for the unconditional vector $x$:
	\begin{proposition}[Matejka and McKay (2015)]\label{MM_result} Let Assumption \ref{No_Model_uncertainty} hold. Then the collection of conditional choice probabilities $P=\{p(\theta)\}_{\theta\in \Theta}$
		solves (\ref{Program1})-(\ref{constrain2}) if and only if the probabilities are given by the generalized MNL formula (\ref{RIlogitprobs}) with the (unconditional) choice probability vector $x$ solving the concave optimization problem
		\begin{equation}\label{Stochastic_Program1}
			\max_{x\in \Delta_n}\EE_{\mu}[H(x, \theta)]
		\end{equation}
		where $H(x,\theta)\triangleq \xi\log (x^\top e^{u(\theta)/\xi})$	and   $\Delta_n\triangleq \Delta(A)$ being the $n$-dimensional simplex.
	\end{proposition}
	Proposition~\ref{MM_result} shows that the prior affects optimal behavior exclusively through the unconditional choice probabilities $x$. This observation is the starting point of our analysis. Once the prior itself becomes ambiguous, robustness can be incorporated by optimizing jointly over the unconditional choice probabilities and the set of feasible priors.
\subsection{A robust approach to RI}

We now relax the benchmark assumption that the decision maker knows the prior distribution over states. Following the maxmin approach to ambiguity of \cite{GILBOA1989141} and \cite{Marinaccietal2006}, we allow the decision maker to evaluate her information strategy against the worst-case prior by solving

	\begin{equation}	\label{Program1_robust_max_min}
		\max_{P\in \mathcal{P}}\min_{\nu\in \mathcal{B}\subseteq
			\Delta(\Theta)}\left\{ \sum_{\theta\in \Theta}	\left[\sum_{a\in A}
		p_{a}(\theta)u(a,\theta)\right]\nu(\theta) -\xi C(P, \nu) \right\} ,
	\end{equation} 
where $\mathcal B\subseteq\Delta(\Theta)$ is a nonempty, compact, and convex ambiguity set, which nests (\ref{Program1}) when $\mathcal{B}=\{\mu\}$. As with the non-robust problem, this has an equivalent unconditional-choice-probability form:
	\begin{proposition}\label{Robust_max_min_equiv} Let $\mathcal B\subseteq\Delta(\Theta)$ be nonempty, compact, and convex. The robust problem  (\ref{Program1_robust_max_min}) is equivalent to
		\begin{equation}	\label{Program1_robust_max_min_2}
			\max_{x\in \Delta_n}\min_{\nu\in \mathcal B}\{\EE_{\nu}(H(x,\theta))\}.
		\end{equation}
	\end{proposition}

	The proof, reported in Appendix~\ref{Proofs}, combines the statewise variational (Donsker--Varadhan) representation of $H(x,\theta)$ underlying the Matejka--McKay characterization with two applications of Sion's (\citeyear{Sion_1958}) minimax theorem: the first interchanges $\max_P$ and $\min_\nu$, using that the objective of (\ref{Program1_robust_max_min}) is concave in $P$ (mutual information is convex in the channel) and convex in $\nu$ (mutual information is concave in the input distribution); the second interchanges $\min_\nu$ and $\max_x$ in the reduced problem, whose objective is concave in $x$ and linear in $\nu$.
	\smallskip

	Proposition~\ref{Robust_max_min_equiv} allows us to work directly with the equivalent unconditional-choice formulation throughout the remainder of the paper, for any nonempty, compact, convex ambiguity set $\mathcal{B}\subseteq\Delta(\Theta)$. In Section~\ref{s3}, we construct such sets using $\phi$-divergences, leading to a tractable formulation that supports the comparative statics and consideration-set analysis developed below.
\section{The DRO-RI Model}\label{s3}

This section develops the Distributionally Robust Rational Inattention (DRO-RI) model. We first construct ambiguity sets for the prior using $\phi$-divergences (\cite{Ali_1966}, \cite{Csizar_1967}, and \cite{LieseVajda1987}).\footnote{The same object is also called an \emph{$f$-divergence} in parts of the literature, following \cite{Csizar_1967}'s original terminology (see also \citealp{Ali_1966}). We adopt the term ``$\phi$-divergence'' to follow the distributionally robust optimization literature (e.g., \citealp{BenTaletal_2013,Shapiro_et_al_2021.Ch7}). The two terms are used interchangeably.} We then show that the resulting max--min problem admits an equivalent finite-dimensional concave formulation, providing a tractable characterization of optimal behavior under prior ambiguity. Finally, we establish the connection between the DRO-RI model and the theory of convex risk measures.
\subsection{$\phi$-divergence functions}

We measure ambiguity using $\phi$-divergences, a broad class of statistical discrepancy measures between probability distributions.

\begin{definition}\label{Phi_Divergence}
	Let $\phi:[0,\infty)\rightarrow\mathbb{R}\cup\{\infty\}$ be a convex function satisfying $\phi(1)=0$. The $\phi$-divergence between two probability vectors $\mu,\nu\in\Delta_m$ is defined by
	\begin{equation}\label{I_phi_divergence}
		I_{\phi}(\nu,\mu)
		=
		\sum_{i=1}^{m}
		\mu_i
		\phi\!\left(\frac{\nu_i}{\mu_i}\right),
	\end{equation}
	where the conventions
	\[
	0\,\phi(a/0)
	=
	a\lim_{t\rightarrow\infty}\frac{\phi(t)}{t},
	\qquad
	0\,\phi(0/0)=0,
	\]
	are adopted.
\end{definition}

\begin{assumption}\label{smooth_divergence}
	Throughout the paper, the divergence function $\phi$ is twice continuously differentiable on $(0,\infty)$ and satisfies
	\[
	\phi(1)=\phi'(1)=0,
	\qquad
	\phi''(1)>0.
	\]
\end{assumption}

	Assumption~\ref{smooth_divergence} is satisfied by most $\phi$-divergences commonly used in statistics and information theory. For example, the choice
	\[
	\phi(t)=t\log t-t+1
	\]
	generates the Kullback--Leibler divergence,
	\[
	I_\phi(\nu,\mu)
	=
	\sum_i
	\nu_i
	\log\!\left(\frac{\nu_i}{\mu_i}\right),
	\]
	for which $\phi''(1)=1$.\footnote{When $\phi$ is differentiable at $t=1$, the function $\varphi(t)=\phi(t)-\phi'(1)(t-1)$ generates an equivalent $\phi$-divergence satisfying $\varphi(1)=\varphi'(1)=0$ and $\varphi\ge0$; see \cite{Pardo2005}.} Table~\ref{table:examples} reports several commonly used $\phi$-divergences together with their associated divergence measures and convex conjugates,
	\[
	\phi^*(s)
	=
	\sup_{t\ge0}\{st-\phi(t)\},
	\]
	which play a central role in the analysis below. Although closed-form expressions for $\phi^*$ are unavailable for some divergences (for example, the $J$-divergence), tractable reformulations remain available through standard tools from convex analysis (\cite{Rockafellar1970}).
	
	{\tiny
		\begin{table}[ht]
			\caption{Some $\phi$-divergence examples and their conjugates}
			\centering
			\resizebox{\hsize}{!}{%
			\begin{tabular}{c c c c c}
				\hline\hline
				Divergence &  $\phi(t)$ &  Explicit form & $I_\phi(\nu,\mu)$  & $\phi^*(s)$\\ [0.5ex]
				\hline
				Kullback-Leibler &$\phi_{kl}(t)$& $t \log t-t+1 $ &$ \sum_{i=1}^m \nu_{i} \log \left(\nu_{i}/\mu_{i}\right)$ & $e^s-1$ \\
				Burg Entropy&$\phi_{b}(t)$ &$  -\log t+t-1 $&$\sum_{i=1}^m \mu_{i} \log \left(\mu_{i}/\nu_{i}\right)$ &$
				-\log (1-s), \quad s<1
				$ \\
				J-Divergence &$\phi_{j}(t)$&$(t-1) \log  t$ &$\sum_{i=1}^m\left(\nu_{i}-\mu_{i}\right) \log \left(\frac{\nu_{i}}{\mu_{i}}\right)$& \text{No closed form} \\
				$\chi^{2}$-distance& $\phi_{c}(t)$ & $\frac{1}{t}(t-1)^{2}$ & $\sum_{i=1}^m \frac{\left(\nu_{i}-\mu_{i}\right)^{2}}{\nu_{i}}$& $2-2 \sqrt{1-s}$ \\
				Modified $\chi^{2}-$ distance & $\phi_{m c}(t)$& ${1\over 2}(t-1)^{2}$ & ${1\over 2}\sum_{i=1}^m\frac{\left(\nu_{i}-\mu_{i}\right)^{2}}{\mu_{i}}$ & $\begin{cases}-{1\over 2} & s<-1 \\ s+s^{2} / 2 & s \geq-1\end{cases}$\\
				Hellinger
				distance & $\phi_{h}(t)$& $(\sqrt{t}-1)^{2}$ & $\sum_{i=1}^m\left(\sqrt{\nu_{i}}-\sqrt{\mu_{i}}\right)^{2}$ & $\frac{s}{1-s}, \quad s<1$\\ 
				$\chi$-divergence
				of $\operatorname{order} k>1$& $\phi_{c a}^{k}(t)$& $|t-1|^{k}$ & $\sum_{i=1}^m \mu_{i}\left|1-\frac{\nu_{i}}{\mu_{i}}\right|^{k}$& $s+(k-1)\left(\frac{|s|}{k}\right)^{k /(k-1)}$\\ 		
				Variation Distance& $\phi_{v}(t)$ & $|t-1|$ & $\sum_{i=1}^m\left|\nu_{i}-\mu_{i}\right|$&  $\begin{cases}-1 & s\leq-1 \\ s  &-1\leq s \leq 1\end{cases}$\\ 	
				Cressie-Read & $\phi_{c r}^{k}(t)$ &$\frac{t^{k}-k t+k-1}{k(k-1)}, k\in\RR \backslash\{0,1\} $& $\frac{1}{k(1 - k)}\left(1-\sum_{i=1}^m \nu_{i}^{k} \mu_{i}^{1-k}\right)$  &  $\frac{1}{k}\left[((k-1) s+1)_{+}^{k_{*}}-1\right]$\\ 	
				& &  & & $k^*={k\over k-1}$\\ 	
				[1ex]
				\hline
			\end{tabular}}
			\label{table:examples}
		\end{table}
	}

\subsection{Uncertainty sets}

We model prior ambiguity by allowing the true prior to lie within a neighborhood of the benchmark prior $\mu$, where proximity is measured by a $\phi$-divergence. Specifically, for a given radius $\rho\ge0$, we define the ambiguity (or uncertainty) set
\begin{equation}\label{Phi_Set}
	\mathcal{B}_\phi(\mu)
	=
	\left\{
	\nu\in\Delta(\Theta):
	I_\phi(\nu,\mu)\le\rho
	\right\}.
\end{equation}
	
	Following \cite{Hansen_Sargent_2001}, the prior $\mu$ represents the DM's benchmark (or \emph{nominal}) distribution, while $\mathcal{B}_\phi(\mu)$ contains all probability distributions whose $\phi$-divergence from $\mu$ does not exceed $\rho$. The parameter $\rho$ therefore measures the degree of robustness: when $\rho=0$, the ambiguity set collapses to the singleton $\{\mu\}$, recovering the benchmark RI model, whereas as $\rho\to\infty$, it expands to the entire simplex $\Delta(\Theta)$, representing maximal ambiguity over the prior.
	\smallskip

	Replacing the singleton prior $\{\mu\}$ in Proposition~\ref{Robust_max_min_equiv} with the ambiguity set $\mathcal{B}_\phi(\mu)$ yields the following distributionally robust optimization problem:
	\begin{equation}\label{DR0-RI-phi}
		\upsilon^\star=\max_{x\in \Delta_n}\min_{\nu\in \mathcal{B}_\phi(\mu)}\{\EE_{\nu}(H(x,\theta))\}.
	\end{equation}
	
  We are now ready to state the main result of this section. Unlike Assumption~\ref{smooth_divergence}, which is used only where the specific differentiable specialization of the worst-case prior is invoked, the duality result below requires only the following minimal regularity, standard in convex duality (\citealp{Rockafellar1970,BenTaletal_2013}).

\begin{assumption}[Duality regularity]\label{Duality_Regularity}
	The divergence-generating function $\phi:[0,\infty)\to\mathbb{R}\cup\{+\infty\}$ is proper, lower semicontinuous, and convex, with $\phi(1)=0$. The benchmark prior $\mu$ has full support (Assumption~\ref{No_Model_uncertainty}).
\end{assumption}

Assumption~\ref{Duality_Regularity} is satisfied by every $\phi$-divergence in Table~\ref{table:examples}, smooth or not, including the (non-smooth) AV$@$R-generating divergence of Remark~\ref{remark_AVaR_smoothness} below. When $\rho>0$, full support of $\mu$ together with $\phi(1)=0$ implies that $\nu=\mu$ is strictly feasible for the constraint $I_\phi(\nu,\mu)\le\rho$ defining $\mathcal B_\phi(\mu)$, i.e. Slater's condition holds, which is all that is needed for the Lagrangian duality argument below; no differentiability of $\phi$ (Assumption~\ref{smooth_divergence}) is required.

\begin{theorem}\label{Thm_DRO-RI-phi}
	Let Assumption~\ref{Duality_Regularity} hold and let $\rho>0$. The DRO-RI problem~(\ref{DR0-RI-phi}) admits the following equivalent finite-dimensional optimization problem:
	\begin{equation}\label{Max_Lag_phi_DRO_phi}
		\max_{x\in\Delta_n,\;\lambda\ge0,\;\eta\in\mathbb{R}}
		\left\{
		-\eta
		-\rho\lambda
		-\lambda
		\EE_{\mu}
		\left[
		\phi^*\!\left(
		\frac{-H(x,\theta)-\eta}{\lambda}
		\right)
		\right]
		\right\}.
	\end{equation}
	Here, $\lambda$ and $\eta$ denote the Lagrange multipliers associated with the ambiguity-set constraint defining $\mathcal{B}_\phi(\mu)$ and the probability constraint $\sum_{\theta\in\Theta}\nu(\theta)=1$, respectively. Moreover, the objective function in (\ref{Max_Lag_phi_DRO_phi}) is jointly concave in $(x,\lambda,\eta)$.
\end{theorem}
	
	Theorem~\ref{Thm_DRO-RI-phi} reduces the original max--min problem to a finite-dimensional concave optimization problem in $(x,\lambda,\eta)$, with the expectation taken under the nominal prior $\mu$. This places the DRO-RI model within the framework of stochastic optimization. When the convex conjugate $\phi^*$ is available in closed form, the problem can be solved directly. Otherwise, standard methods such as \emph{Sample Average Approximation} (SAA) (\cite{Shapiro_et_al_2021.ch8}) provide tractable numerical approximations.\footnote{To apply SAA, draw an i.i.d.\ sample $\theta^1,\ldots,\theta^T$ from $\mu$, and solve
		\begin{equation*}
			\max_{x\in\Delta_n,\;\lambda\ge0,\;\eta}
			\left\{
			-\eta-\rho\lambda
			-\frac{\lambda}{T}
			\sum_{t=1}^{T}
			\phi^{*}\!\left(
			\frac{-H(x,\theta^t)-\eta}{\lambda}
			\right)
			\right\}.
	\end{equation*}}

\begin{proposition}\label{Computing_nu}
	Let Assumption~\ref{Duality_Regularity} hold, and let $(x^\star,\lambda^\star,\eta^\star)$ be an optimal solution to problem~(\ref{Max_Lag_phi_DRO_phi}), and define
	\[
	s^\star(\theta)
	\triangleq
	-\frac{H(x^\star,\theta)+\eta^\star}{\lambda^\star}.
	\]
	For $\lambda^\star>0$, the worst-case prior $\nu^\star$ is characterized by
	\begin{align}
		\frac{\nu^\star(\theta)}{\mu(\theta)}
		&\in
		\partial\phi^*\!\left(s^\star(\theta)\right),
		\label{RP_1}\\
		\sum_{\theta\in\Theta}
		\mu(\theta)
		\phi\!\left(
		\frac{\nu^\star(\theta)}{\mu(\theta)}
		\right)
		&=
		\rho,
		\label{RP_2}\\
		\sum_{\theta\in\Theta}
		\nu^\star(\theta)
		&=
		1.
		\label{RP_3}
	\end{align}
\end{proposition}
	
	Proposition~\ref{Computing_nu} follows from the first-order conditions of problem~(\ref{Max_Lag_phi_DRO_phi}). Equation~(\ref{RP_1}) characterizes the worst-case prior through Fenchel's equality,
	\[
	\frac{\nu^\star(\theta)}{\mu(\theta)}
	\in
	\partial\phi^*(s^\star(\theta))
	\iff
	s^\star(\theta)
	\in
	\partial\phi\!\left(
	\frac{\nu^\star(\theta)}{\mu(\theta)}
	\right),
	\]
	which reduces to
	\[
	\nu^\star(\theta)
	=
	\mu(\theta)\phi^{*\prime}(s^\star(\theta))
	\]
	whenever $\phi^*$ is differentiable at $s^\star(\theta)$.\footnote{$\phi^*$ is proper, closed, and convex whenever $\phi$ is (\citealp[Thm.~12.2]{Rockafellar1970}), and is therefore subdifferentiable on the relative interior of its domain (\citealp[Thm.~23.4]{Rockafellar1970}).} Equation~(\ref{RP_2}) states that the ambiguity-set constraint binds at the optimum by complementary slackness, while (\ref{RP_3}) is the probability normalization condition.
	\begin{corollary}
		\label{optimized_value}
		Let Assumption~\ref{Duality_Regularity} hold, and suppose that $\phi^\ast$ is differentiable at
		$s^\star(\theta)$ for every $\theta\in\Theta$. Let
		$(x^\star,\lambda^\star,\eta^\star)$ be an optimal solution to problem~(\ref{Max_Lag_phi_DRO_phi})
		with $\lambda^\star>0$. Then
		\[
		\upsilon^\star
		=
		E_{\nu^\star}[H(x^\star,\theta)],
		\]
		where
		\[
		\nu^\star(\theta)
		=
		\mu(\theta)
		\phi^{*\prime}
		\left(
		-\frac{H(x^\star,\theta)+\eta^\star}{\lambda^\star}
		\right),
		\qquad
		\theta\in\Theta.
		\]
	\end{corollary}

	\begin{example}\label{Kullback_Leibler}
		For the Kullback--Leibler divergence,
		\[
		\phi_{kl}^{*\prime}(s)=e^s.
		\]
		Hence, equations~(\ref{RP_1})--(\ref{RP_3}) admit the closed-form solution
		\begin{equation}\label{KL_Expression}
			\nu^\star(\theta)
			=
			\frac{\mu(\theta)e^{-H(x^\star,\theta)/\lambda^\star}}
			{\EE_\mu\!\left[e^{-H(x^\star,\theta)/\lambda^\star}\right]},
			\qquad
			\forall\,\theta\in\Theta,
		\end{equation}
		which is the familiar Gibbs (or exponential-tilting) representation of the worst-case prior.\footnote{From (\ref{RP_1}),
			\[
			\nu^\star(\theta)
			=
			\mu(\theta)
			e^{-(H(x^\star,\theta)+\eta^\star)/\lambda^\star}.
			\]
			Summing over $\theta$ and imposing (\ref{RP_3}) yields
			\[
			e^{\eta^\star/\lambda^\star}
			=
			\EE_\mu\!\left[
			e^{-H(x^\star,\theta)/\lambda^\star}
			\right].
			\]
			Substituting this expression back into the previous equation gives (\ref{KL_Expression}).}
		This exponential-tilting representation coincides with the worst-case prior in entropy-based distributionally robust optimization and provides the benchmark case for the more general $\phi$-divergence formulations developed below.
	\end{example}
	
	\subsection{DRO-RI and risk measures}
		The DRO-RI model admits a natural interpretation through the theory of monetary utility functionals and risk measures under ambiguity (\citealp{FollmerSchied2002,Shapiro_et_al_2013}). For a fixed unconditional choice probability vector $x\in\Delta_n$, define the \emph{concave (distributionally robust) utility functional}
	\begin{equation}\label{Concave_Utility}
		u_\phi(x)\;\triangleq\;\min_{\nu\in\mathcal{B}_\phi(\mu)}\EE_\nu[H(x,\theta)],
	\end{equation}
	which is exactly the inner minimization in (\ref{DR0-RI-phi}), so that Theorem~\ref{Thm_DRO-RI-phi} characterizes $\upsilon^\star=\max_{x\in\Delta_n}u_\phi(x)$. Viewed as a functional of the payoff profile $H(x,\cdot)$, $u_\phi$ is concave (a pointwise infimum of the functionals $\EE_\nu[\cdot]$, each linear in $H(x,\cdot)$), monotone ($H_1\geq H_2$ pointwise implies the value of $u_\phi$ at $H_1$ is at least its value at $H_2$, since every $\nu\in\mathcal{B}_\phi(\mu)$ is a probability measure), and cash-additive (replacing $H$ by $H+c$ for a constant $c$ adds $c$ to $u_\phi$, again because $\sum_\theta\nu(\theta)=1$). These three properties are precisely the defining axioms of a \emph{concave monetary utility function} in the sense of \cite{FollmerSchied2002}, the dual object to the convex risk measures conventionally used to price losses: $u_\phi(x)=-\rho_\phi(-H(x,\cdot))$ for $\rho_\phi(Z)\triangleq\sup_{\nu\in\mathcal{B}_\phi(\mu)}\EE_\nu[Z]$. It is this convex risk measure $\rho_\phi$, evaluated at the loss $\tilde H\triangleq-H$, that coincides with the functional $\mathcal{R}$ used below and throughout Section~\ref{s44}: $\mathcal{R}(x)=\rho_\phi(\tilde H(x,\cdot))=-u_\phi(x)$.

	The functional $u_\phi$ also situates the DRO-RI model within the theory of ambiguity-sensitive preferences. Its penalized counterpart, $\min_{\nu\in\Delta(\Theta)}\{\EE_\nu[H(x,\theta)]+\zeta I_{\phi}(\nu,\mu)\}$ --- generalized to a second, independent divergence in \S\ref{s34}--\S\ref{s36} --- is a variational-preference functional in the sense of \cite{Marinaccietal2006}, with ambiguity index $c(\nu)=\zeta I_\phi(\nu,\mu)$; and $u_\phi$ itself reduces to the Gilboa--Schmeidler (\citeyear{GILBOA1989141}) maxmin expected utility criterion when $\mathcal{B}_\phi(\mu)$ is treated as an unpenalized set of priors rather than a $\phi$-divergence ball. The optimizer $\nu^\star(x)$ in (\ref{Concave_Utility}) plays the role of the worst-case probability measure studied by \cite{Cuchieroetal2022}; their analysis of when robust risk measures admit a classical single-reference-measure representation provides a natural complement to our results.

	For continuity with the risk-measure literature, which conventionally states Average Value-at-Risk and entropic risk as functionals of a loss rather than a gain, we work for the remainder of this subsection --- and throughout Section~\ref{s44} --- with the convex risk measure $\mathcal{R}=\rho_\phi\circ\tilde H=-u_\phi$; every displayed $-\mathcal{R}(x)$ below may equivalently be read as the concave utility $u_\phi(x)$ evaluated directly on the gain $H(x,\cdot)$, with no sign convention to track. The connection becomes particularly transparent for two important choices of the divergence function. We begin with Average Value-at-Risk (AV$@$R). To avoid a clash with the exponent $\alpha$ used in Section~\ref{s44}, we denote the AV$@$R confidence level by $\beta$. For $\beta\in(0,1]$, define
	$$
	\phi(t)=\begin{cases}
		0, & \text{ $0\leq t\leq 1/\beta$,}\\
		+\infty & \text{otherwise}.
	\end{cases}
	$$
	
	The associated uncertainty set is given by:
	\begin{equation}\label{Uncertainty_CVaR}
		\mathcal{B}_{\phi}(\mu)=\left\{\nu \in \Delta(\Theta): {\nu(\theta)\over \mu(\theta)}\in [0,\beta^{-1}]\right\},
	\end{equation}
	which does not depend on the radius $\rho$. In this case the conjugate of $\phi$ is  given by $\phi^{*}(s)=\max \left\{0, \beta^{-1} s\right\}\triangleq\left[\beta^{-1} s\right]_{+}$.
	
	 Since $\lambda\phi=\phi$ for every $\lambda>0$, for a fixed  $x\in\Delta_n$, we define  the functional $\R(x)$ as follows:
	\begin{eqnarray}\label{concentration_ex1}
		\R(x)&=&\inf _{\eta\in\RR, \lambda \geq 0}\left\{\lambda \rho+\eta+\beta^{-1} \mathbb{E}_{\mu}[\tilde{H}(x,\theta)-\eta]_{+}\right\}\\
		&=&\inf _{\eta\in \RR}\left\{\eta+\beta^{-1} \mathbb{E}_{\mu}[\tilde{H}(x,\theta)-\eta]_{+}\right\}\nonumber 
	\end{eqnarray}
	where $\tilde{H}(x,\theta)\triangleq -H(x,\theta)$ is a convex function on $x$.  
	
	Expression~(\ref{concentration_ex1}) is the AV$@$R risk measure at confidence level
	$\beta$. Hence, Theorem~\ref{Thm_DRO-RI-phi} becomes
	\begin{equation}\label{R_VaR}
	\max_{x\in\Delta_n}\{-\mathcal{R}(x)\},
	\end{equation}
	so the robust decision maker chooses $x$ to minimize the AV$@$R of the loss
	$\widetilde H(x,\theta)=-H(x,\theta)$.

	\begin{remark}[A non-smooth $\phi$]\label{remark_AVaR_smoothness} The AV$@$R-generating $\phi$ above --- $0$ on $[0,1/\beta]$, $+\infty$ elsewhere --- has a kink at $t=1/\beta$ and is therefore not twice continuously differentiable: it fails Assumption \ref{smooth_divergence}, but satisfies the strictly weaker Assumption~\ref{Duality_Regularity} (it is proper, lower semicontinuous, and convex, with $\phi(1)=0$ since $\beta\in(0,1]$ implies $1\in[0,1/\beta]$), under which Theorem \ref{Thm_DRO-RI-phi} is stated. This is a standard, well-behaved exception in the distributionally robust optimization literature (e.g. \cite{Kuhn_et_al_2019}), and Theorem \ref{Thm_DRO-RI-phi}'s Fenchel--Rockafellar duality argument goes through for any $\phi$ satisfying Assumption~\ref{Duality_Regularity}, smooth or not, since it only requires the existence of the conjugate $\phi^*$, not its differentiability. Differentiability is used elsewhere, however: Proposition \ref{Robust_Consideration_Sets} requires $\phi^*$ to be differentiable at the optimal arguments $s^\star(\theta)$, which can fail here --- $\phi^{*\prime}(s)=0$ for $s<0$ and $1/\beta$ for $s>0$, with the derivative undefined exactly at the kink $s=0$, which is where states tied at the V$@$R threshold sit. This is not a technical obstruction so much as the source of the economically interesting feature of the AV$@$R case: the worst-case prior $\nu$ need not have full support on $\Theta$, and may be set-valued at ties. Corollary \ref{Corollary_AVaR_DRCS} below adapts the proof of Proposition \ref{Robust_Consideration_Sets} directly to this case using subgradients rather than derivatives.
	\end{remark}
	\smallskip

Second, consider the Kullback--Leibler divergence,
\[
\phi_{kl}(t)=t\log t-t+1.
\]
Since
\[
(\lambda\phi)^*(s)
=
\lambda\left(e^{s/\lambda}-1\right),
\]
Theorem~\ref{Thm_DRO-RI-phi} yields
\begin{equation}\label{concetration_ex2}
	\R_\rho(x)
	=
	\inf_{\lambda\ge0,\;\eta\in\mathbb R}
	\left\{
	\lambda\rho+\eta
	+\lambda e^{-\eta/\lambda}
	\EE_\mu
	\!\left[
	e^{\tilde H(x,\theta)/\lambda}
	\right]
	-\lambda
	\right\}.
\end{equation}

The first-order condition with respect to $\eta$ is $1-e^{-\eta/\lambda}\EE_\mu[e^{\tilde H(x,\theta)/\lambda}]=0$, so that $\eta^*=\lambda\log\EE_{\mu}[e^{\tilde{H}(x,\theta)/\lambda}]$. Substituting back, we have
\begin{equation}\label{concentration_ex22}
	\R_\rho(x)
	=
	\inf_{\lambda>0}
	\left\{
	\lambda\rho
	+
	\lambda
	\log
	\EE_\mu
	\left[
	e^{\tilde H(x,\theta)/\lambda}
	\right]
	\right\}.
\end{equation}

At $\rho=0$, the ambiguity set $\mathcal B_\phi(\mu)$ collapses to the singleton $\{\mu\}$, so there is no robustness left to price: the infimum in (\ref{concentration_ex22}) is approached as $\lambda\to\infty$, and since $\lambda\log\EE_\mu[e^{\tilde H(x,\theta)/\lambda}]\to\EE_\mu[\tilde H(x,\theta)]$ as $\lambda\to\infty$ (a first-order expansion of the log around $1$), we have $\R_0(x)=\EE_\mu[\tilde H(x,\theta)]$, the ordinary (non-robust) expected loss, not the entropic risk measure.

The entropic-risk interpretation instead applies to the \emph{multiplier} (penalty) formulation at a fixed, exogenously specified $\lambda>0$, in which deviations of $\nu$ from $\mu$ are penalized rather than constrained to a $\rho$-ball:
\[
\lambda\log\EE_\mu\!\left[e^{\tilde H(x,\theta)/\lambda}\right],
\]
which is exactly the entropic risk measure at risk-aversion parameter $\lambda$ (\cite{Shapiro_et_al_2013}). The constrained (KL-ball) model $\R_\rho(x)$ in (\ref{concentration_ex22}) and this multiplier model are related through Lagrangian duality rather than through the special case $\rho=0$: for every $\rho>0$ there is an optimal dual multiplier $\lambda^\star(\rho)>0$, the shadow price of the divergence-radius constraint, such that $\R_\rho(x)=\lambda^\star(\rho)\rho+\lambda^\star(\rho)\log\EE_\mu[e^{\tilde H(x,\theta)/\lambda^\star(\rho)}]$; the constrained and multiplier problems therefore agree once $\lambda$ is set equal to the shadow price associated with $\rho$, not by setting $\rho=0$.

In this case, Shannon mutual information governs both the attention cost $\xi C(P,\mu)$ and the ambiguity set $\mathcal B_\phi(\mu)$, giving rise to what we call the \emph{double Shannon} model. Through the multiplier correspondence just described, this coincides with the model of \cite{Hansen_et_al_2022} at $\lambda=\lambda^\star(\rho)$, which therefore admits an equivalent interpretation as an entropic-risk minimization problem. Equivalently, in the concave-utility language of (\ref{Concave_Utility}), the double Shannon model maximizes the entropic-robust utility $u_{\phi_{kl}}(x)=-\lambda\log\EE_\mu\!\left[e^{-H(x,\theta)/\lambda}\right]$ directly, with no loss-based sign convention to track.

	\subsection{DRO-RI and  tail performance} We next consider the Cressie--Read family of divergences
	(\cite{CressieRead1984}), connecting program
	(\ref{Max_Lag_phi_DRO_phi}) to an explicit expression for the convex
	conjugate $\phi_{cr}^{*k}$.
	\begin{equation}\label{Cressie_Read_Divergences}
		\phi_{cr}^k(t)
		\triangleq
		\frac{t^k-kt+k-1}{k(k-1)},
	\end{equation}
	whose convex conjugate is
	\begin{equation}\label{Cressie_Read_Conjugate}
		\phi_{cr}^{*k}(s)
		\triangleq
		\frac1k
		\left(
		\bigl((k-1)s+1\bigr)_+^{\,k_*}
		-
		1
		\right).
	\end{equation}
	
	For $t<0$, define $\phi_{cr}^k(t)=+\infty$. The expression in (\ref{Cressie_Read_Divergences}) is valid for $k\neq0,1$; the limiting cases $k=0$ and $k=1$ are defined by continuity. The
	Cressie--Read family (\ref{Cressie_Read_Divergences}) includes the
	modified $\chi^2$-divergence ($k=2$), one half of the $\chi^2$-distance of Table~\ref{table:examples} ($k=-1$), the empirical likelihood divergence
	\[
	\phi_{cr}^{0}(t)=-\log t+t-1,
	\]
	and the Kullback--Leibler divergence
	\[
	\phi_{cr}^{1}(t)=t\log t-t+1
	\]
	as special cases.
	\smallskip
	
	We now characterize the DRO-RI problem for the Cressie--Read family.
	\begin{proposition}\label{Simplified_Dual1}For any prior  $\mu$ on $\Delta(\Theta)$, $k \in(1, \infty), k_{*}=k /(k-1)$, any $\rho>0$, and $m_{k}(\rho):=(1+k(k-1) \rho)^{\frac{1}{k}}$, we have for all $x \in \Delta_n$,
		\begin{equation}\label{Simplified_Dual}
			\mathcal{R}_{k}(x;\rho)=\inf _{\eta \in \mathbb{R}}\left\{m_{k}(\rho) \mathbb{E}_{\mu}\left[(\tilde{H}(x , \theta)-\eta)_{+}^{k_{*}}\right]^{\frac{1}{k_{*}}}+\eta\right\}.
		\end{equation}
	
	Furthermore, the DRO-RI problem (\ref{Max_Lag_phi_DRO_phi}) is equivalently written as
		\begin{equation}\label{Sol_CR_family}
			\max_{x\in\Delta_n}\{-\mathcal{R}_k(x;\rho)\}.
		\end{equation}
	\end{proposition}
	
	The simplified dual~(\ref{Simplified_Dual}) gives a direct interpretation of robustness under the Cressie--Read family. For a fixed $x$, the functional $\mathcal{R}_k(x;\rho)$ depends only on losses exceeding the endogenous threshold $\eta^\star(x)$ through their $L^{k_*}(\mu)$-norm. Since $k_{*}>1$, this norm places relatively greater weight on large realizations of $\widetilde H(x,\theta)$ --- that is, on large losses --- thereby emphasizing states in which the decision problem performs poorly. In this sense, the Cressie--Read specification induces a form of tail-sensitive robustness.
	
	The parameter $k_{*}$ controls the degree of this tail sensitivity. As $k$ increases, $k_{*}=k/(k-1)$ decreases toward one, so the $L^{k_{*}}(\mu)$-norm becomes progressively less sensitive to extreme losses. The range $k\in(1,\infty)$ is therefore particularly convenient when $\widetilde H(x,\theta)$ has finite $k_{*}$-moments under $\mu$ and provides a tractable family for studying how robustness varies with sensitivity to adverse states.

\section{ Composite and Nested Ambiguity}\label{s45}

This section extends the baseline DRO-RI model of Section~\ref{s3} to allow separate, independently calibrated sources of robustness. Subsection~\ref{s34} lets the attention technology (cost function) and the ambiguity set over priors be governed by two different $\phi$-divergences, $\phi_1$ and $\phi_2$, and shows that the resulting composite model remains tractable. Subsection~\ref{s36} instead keeps the attention technology fixed and adds a second, outer layer of ambiguity around the reference prior itself.

	\subsection{Generalized attention costs: a composite robustness approach}\label{s34}
	
	Thus far, information acquisition has been modeled using the Shannon mutual information cost $C(P,\mu)$ in (\ref{Mutual_information_cost}). The corresponding \emph{statewise} value function admits the variational representation
	
	\begin{equation}\label{H_as_KL_variational}
		H(x,\theta)=\max_{p(\theta)\in \Delta_n}\left\{\sum_{a\in A}p_a(\theta)u(a,\theta)-\xi\, I_{\phi_{kl}}\big(p(\theta)\,\|\,x\big)\right\}.
	\end{equation}

This representation, which is also used in the proof of Proposition~\ref{Robust_max_min_equiv} in Appendix~\ref{Proofs}, makes clear that Shannon mutual information enters the RI problem through a Kullback--Leibler penalty between the conditional choice probabilities $p(\theta)$ and the unconditional choice probabilities $x$.

	Expression~(\ref{H_as_KL_variational}) reveals that the Shannon-cost RI model already embeds a $\phi$-divergence, namely $\phi_{kl}$, at the level of the \emph{attention technology}, with $x$ rather than $\mu$ serving as the reference distribution. Motivated by the composite two-divergence representation of robust risk measures in \cite{Jinetal2024}, building on the optimized certainty equivalent of \cite{BenTalTeboulle2007}, we therefore allow both the attention technology and the prior to be governed by $\phi$-divergences, which need not coincide. This yields a composite robustness model that contains the penalized form of the DRO-RI specification of Section~\ref{s3} as a special case and, consequently, also encompasses the model of \cite{Hansen_et_al_2022}.
		\smallskip

	\subsubsection{Generalized attention costs}
	
	Let $\phi_1$ satisfy Assumption~\ref{smooth_divergence}. For a conditional choice rule $p(\theta)\in\Delta_n$ and an unconditional choice probability vector $x\in\Delta_n$, define the \emph{generalized attention cost}
	\begin{equation}\label{Generalized_Cost}
		C_{\phi_1}(p(\theta),x)\triangleq \sum_{a\in A}x_a\,\phi_1\!\left(\frac{p_a(\theta)}{x_a}\right).
	\end{equation}
	When $\phi_1=\phi_{kl}$, (\ref{Generalized_Cost}) reduces to the Shannon cost of Section~\ref{s2}. In that case, the optimal reference distribution $x$ coincides with the induced marginal choice probabilities,
	$$
	q(a)=\sum_{\theta\in\Theta}p_a(\theta)\mu(\theta),
	$$
	through the Matejka--McKay equivalence underlying Proposition~\ref{Robust_max_min_equiv}. This fixed-point relation relies on the particular structure of Shannon entropy and need not extend to a general $\phi_1$.
	
	For $\phi_1\neq\phi_{kl}$, we therefore interpret $x$ more generally as a reference distribution for conditional choice probabilities rather than imposing that it coincide with the induced marginal. This interpretation is closely related to the use of reference choice probabilities in generalized information-cost models such as \cite{Fosgerauetal2020} and \cite{Caplin_et_al2022}, neither of which considers robustness with respect to the prior. Paralleling (\ref{Stochastic_Program1}), define the \emph{generalized indirect utility}
	
		\begin{equation}\label{Generalized_H}
		H_{\phi_1}(x,\theta)\triangleq \max_{p(\theta)\in \Delta_n}\left\{\sum_{a\in A}p_a(\theta)u(a,\theta)-\xi\, C_{\phi_1}(p(\theta),x)\right\}.
	\end{equation}
	The next lemma adapts the Ben--Tal--Teboulle representation of optimized certainty equivalents (\citealp{BenTalTeboulle2007}; see also Section 2 and Theorem 1 of \citealp{Jinetal2024}) to our discrete-action environment, and provides a dual (closed-form) expression for $H_{\phi_1}$.
\begin{lemma}\label{Generalized_H_Dual}
Let Assumption \ref{smooth_divergence} hold for $\phi_1$. Then, for all $x\in\Delta_n$ and $\theta\in\Theta$,
\begin{equation}\label{Generalized_H_Dual_Eq}
H_{\phi_1}(x,\theta)
=
\inf_{\eta_\theta\in\mathbb{R}}
\left\{
\eta_\theta
+
\xi\sum_{a\in A}x_a\,\phi_1^*\!\left(
\frac{u(a,\theta)-\eta_\theta}{\xi}
\right)
\right\}.
\end{equation}
When $\phi_1=\phi_{kl}$, so that $\phi_1^*(s)=e^s-1$, the minimization over $\eta_\theta$ can be solved in closed form, yielding
\[
H_{\phi_1}(x,\theta)
=
H(x,\theta)
=
\xi\log\left(
\sum_{a\in A}x_a e^{u(a,\theta)/\xi}
\right).
\]
\end{lemma}
	\smallskip
	
	\subsubsection{Composite DRO-RI: a penalty (multiplier) formulation} We now robustify (\ref{Generalized_H}) with respect to prior ambiguity as well, following \cite{Hansen_Sargent_2001} and matching the multiplier formulation of \cite{Jinetal2024} (eq.~(6)) and \cite{Hansen_et_al_2022} (Problem 2): we penalize deviations of $\nu$ from $\mu$ directly rather than constraining them to a $\rho$-ball, replacing $(\rho,\lambda)$ with a single penalty weight $\zeta>0$. Define the \emph{composite} DRO-RI problem
	\begin{equation}\label{Composite_DRO-RI}
		\upsilon^\star_{\phi_1,\phi_2}(\zeta)=\max_{x\in \Delta_n}\min_{\nu\in \Delta(\Theta)}\left\{\EE_{\nu}\big(H_{\phi_1}(x,\theta)\big)+\zeta\, I_{\phi_2}(\nu,\mu)\right\},
	\end{equation}
	for a general $\phi_2$-divergence, where $\zeta$ penalizes the outer prior-ambiguity divergence exactly as $\xi$ penalizes the inner attention-cost divergence in (\ref{Generalized_Cost}), both layers penalty-based as in \cite{Jinetal2024}. Problem (\ref{Composite_DRO-RI}) nests (i) the penalized (multiplier) counterpart of the DRO-RI model of \S \ref{s3} when $\phi_1=\phi_{kl}$, and (ii) \cite{Hansen_et_al_2022}'s robust Shannon-RI model when $\phi_1=\phi_2=\phi_{kl}$.
	
	\begin{theorem}[Composite DRO-RI]\label{Composite_Thm} Let Assumption \ref{smooth_divergence} hold for $\phi_1$ and $\phi_2$. The composite DRO-RI problem (\ref{Composite_DRO-RI}) is equivalent to
		\begin{equation}\label{Composite_Dual}
			\upsilon^\star_{\phi_1,\phi_2}(\zeta)=\max_{x\in \Delta_n,\,\eta\in\RR}\left\{\eta-\zeta\,\EE_{\mu}\!\left[\phi_2^{*}\!\left(\frac{\eta-H_{\phi_1}(x,\theta)}{\zeta}\right)\right]\right\},
		\end{equation}
		where $H_{\phi_1}(x,\theta)$ is given in closed (nested) form by Lemma \ref{Generalized_H_Dual}. The objective in (\ref{Composite_Dual}) is jointly concave in $(x,\eta)$, and an optimal $(x^*,\eta^*)$, together with the state-wise optimizers $\eta_\theta^*(x^*)$ of Lemma \ref{Generalized_H_Dual}, jointly recovers the robust attention policy $x^*$, the optimal conditional choice rule $p^\star(\theta)$, and the worst-case prior $\nu^*$.
	\end{theorem}
	The proof in Appendix~\ref{Proofs} dualizes the normalization constraint $\sum_{\theta}\nu(\theta)=1$, treating $H_{\phi_1}(x,\theta)$ as a fixed statewise value function, as in the proof of Theorem~\ref{Thm_DRO-RI-phi}.
	By Lemma~\ref{Generalized_H_Dual}, $H_{\phi_1}(x,\theta)$ is concave in
	$x$. The resulting representation is therefore naturally nested: the outer problem is a concave maximization over $(x,\eta)$, while each
	evaluation of its objective invokes the state-by-state minimization over $\eta_\theta$ characterized in Lemma~\ref{Generalized_H_Dual}.
	
	We retain this nested formulation rather than rewriting the problem as a single optimization over $(x,\eta,{\eta_\theta})$. Such a reformulation
	would require an additional minimax interchange that is not needed for our results. The nested representation in (\ref{Composite_Dual}) is
	therefore both sufficient for the analysis and directly amenable to computation.
		\smallskip
	
	One might instead try to apply \cite[Thm.1]{Jinetal2024} directly by embedding $(\Theta,A)$ into the product space and defining
	$$
	P_0(\theta,a)=\mu(\theta)x_a,\qquad
	Q(\theta,a)=\nu(\theta)x_a,\qquad
	\bar Q(\theta,a)=\nu(\theta)p_a(\theta).$$
	
	This construction reproduces the two divergence terms appearing in our model. However, Theorem 1 of \cite{Jinetal2024} dualizes a single global normalization through two scalar multipliers, whereas our conditional choice rule requires a separate normalization constraint
	$$
	\sum_{a\in A}p_a(\theta)=1$$
	
	for every $\theta\in\Theta$. Because the decision maker's strategy is state contingent, these statewise constraints cannot be absorbed into the same static dual representation without introducing additional multipliers and an accompanying minimax argument.
	
	Theorem~\ref{Composite_Thm} should therefore be viewed as a state-indexed, nested extension of the static composite representation in \cite{Jinetal2024}. The distinction reflects the economic structure of the problem: our object is a conditional attention technology rather than a one-shot risk evaluation.
	
		\smallskip
	
	Finally, the consideration-set characterization of Section~\ref{s44} extends naturally to the composite model. In Proposition~\ref{Robust_Consideration_Sets}, replace $H(x^\star,\theta)$ by $H_{\phi_1}(x^\star,\theta)$ and the Shannon conditional choice probabilities in (\ref{RIlogitprobs}) by
	$$
	p_a^\star(\theta)= x_a^\star\,
	\phi_1^{*\prime}\left(
	\frac{u(a,\theta)-\eta_\theta^\star}{\xi}
	\right),$$
	
	where $\eta_\theta^\star$ is the statewise optimizer characterized in Lemma~\ref{Generalized_H_Dual}. Under the maintained differentiability conditions, the same first-order argument then characterizes the robust consideration set $A(x^\star)$ under a general, possibly non-Shannon, attention technology. Section~\ref{s52} specializes this characterization to the Cressie--Read case $\phi_1=\phi_{cr}^k$.
	
	\subsection{Robustifying the reference prior itself}\label{s36} The composite model of \S \ref{s34} generalizes the attention layer, leaving prior ambiguity as a single $\phi_2$-divergence around a baseline. We now generalize the \emph{prior} layer instead: \cite{Hansen_et_al_2022} robustify $\mu$ against a baseline $\hat\mu$ via relative entropy, but $\hat\mu$ itself --- typically an estimate --- need not be trusted either. We add a second, outer layer of ambiguity around $\hat\mu$ using a general $\phi_2$-divergence (the machinery of Theorem \ref{Thm_DRO-RI-phi}, via \citealp{BenTaletal_2013}), so that \cite{Hansen_et_al_2022}'s baseline is itself treated as distributionally robust.
	\smallskip

	This two-layer structure is closely related to the decision criterion axiomatized by \cite{CerreiaVioglioetal2026}, who distinguish a set $Q$ of \emph{structured} models from \emph{unstructured} alternatives and evaluate acts $f$ by
	$$\min_{q\in Q}\min_{p\in\Delta}\{\EE_p[u(f)]+\lambda\, I_{\phi_{kl}}(p,q)\}.$$ In their terminology, $\hat\mu$ plays the role of a structured model and $\mu$ that of an unstructured one. Problem (\ref{Nested_Problem}) below replaces their hard restriction $\hat\mu\in Q$ with the soft penalty $\zeta\, I_{\phi_2}(\hat\mu,\mu_0)$; their criterion (with $H(x,\theta)$ in place of $u(f)$) is recovered when this penalty is replaced by the convex indicator of $Q$ (equal to $0$ on $Q$ and $+\infty$ otherwise). In contrast to \cite{CerreiaVioglioetal2026}, whose decision maker is fully informed, our decision maker chooses her information structure, so misspecification concerns also shape what she attends to.
	\smallskip
	
	Following \cite{Hansen_Sargent_2001} and \cite{Hansen_et_al_2022}'s Problem 2, we use the multiplier (penalty) form for both layers. For penalty weights $\psi,\zeta>0$ and a deeper reference prior $\mu_0$, define
	\begin{equation}\label{Nested_Problem}
		J(\mu_0;\psi,\zeta)=\max_{x\in\Delta_n}\ \min_{\hat\mu\in\Delta(\Theta)}\ \min_{\mu\in\Delta(\Theta)}\left\{\EE_\mu\big(H(x,\theta)\big)+\psi\, I_{\phi_{kl}}(\mu,\hat\mu)+\zeta\, I_{\phi_2}(\hat\mu,\mu_0)\right\},
	\end{equation}
	for a general $\phi_2$-divergence.

	Since both inner operators are minimizations over probability distributions, combining them requires no additional minimax argument. In particular,
	$\min_{\hat\mu}\min_{\mu}{\cdot}=\min_{(\hat\mu,\mu)}{\cdot},$
	so the two layers can be treated as a single minimization over the joint pair $(\hat\mu,\mu)$.
\begin{theorem}[Nested prior robustness]\label{Nested_Thm} Let Assumption \ref{smooth_divergence} hold for $\phi_2$. Problem (\ref{Nested_Problem}) is equivalent to
	\begin{equation}\label{Nested_Dual}
		J(\mu_0;\psi,\zeta)=\max_{x\in\Delta_n,\,\eta_1,\eta_2\in\RR}\left\{\eta_1+\eta_2+\psi-\zeta\,\EE_{\mu_0}\!\left[\phi_2^{*}\!\left(\frac{\eta_2+\psi\,e^{(\eta_1-H(x,\theta))/\psi}}{\zeta}\right)\right]\right\}.
	\end{equation}
	Moreover: 
	\begin{enumerate}
	\item[(i)] as $\zeta\to\infty$ (so that the worst-case $\hat\mu^\star\to\mu_0$, rather than $\hat\mu=\mu_0$ holding at any finite $\zeta$), (\ref{Nested_Dual}) collapses to $\max_x\{-\psi\log\EE_{\mu_0}[e^{-H(x,\theta)/\psi}]\}$, exactly \cite{Hansen_et_al_2022}'s Problem 2/eq.~(11).
	\item[(ii)] as $\psi\to\infty$ (so that the worst-case $\mu^\star-\hat\mu^\star\to0$, rather than $\mu=\hat\mu$ holding at any finite $\psi$), (\ref{Nested_Dual}) collapses to $\max_{x,\eta}\{\eta-\zeta\EE_{\mu_0}[\phi_2^*((\eta-H(x,\theta))/\zeta)]\}$, exactly the (single-layer) penalty form (\ref{Composite_Dual}) of Theorem \ref{Composite_Thm} with $\phi_1=\phi_{kl}$ (so that $H_{\phi_1}=H$) and nominal prior $\mu_0$.
	\end{enumerate}
\end{theorem}

The proof in Appendix~\ref{Proofs} dualizes the two normalization constraints jointly and then minimizes pointwise over $(\mu(\theta),\hat\mu(\theta))$. The minimization is carried out first over $\mu(\theta)$ in closed form, yielding the Gibbs step familiar from \cite{Hansen_et_al_2022} and Example~\ref{Kullback_Leibler}, and then over $\hat\mu(\theta)$ through the conjugate $\phi_2^*$. No minimax interchange is required at any stage of the argument.
	\smallskip
	
	Theorem~\ref{Nested_Thm} shows that allowing the baseline prior in \cite{Hansen_et_al_2022} to be itself ambiguous does not sacrifice tractability. The representation in (\ref{Nested_Dual}) remains a finite-dimensional concave program in $(x,\eta_1,\eta_2)$ for any $\phi_2$ satisfying the maintained assumptions, while the model of \cite{Hansen_et_al_2022} and the penalized single-layer DRO-RI model of Theorem \ref{Composite_Thm} (with $\phi_1=\phi_{kl}$) are recovered as limiting cases in which one of the two prior-ambiguity layers vanishes. Unlike the extension in Section~\ref{s34}, the present construction leaves the Shannon attention technology unchanged. Hence the unconditional choice probabilities retain their standard Matejka--McKay interpretation, while both robustness layers operate exclusively over $\Delta(\Theta)$, as in the baseline model of Section~\ref{s3}.
	\section{Robust Consideration Sets} \label{s44}
	We use Theorem~\ref{Thm_DRO-RI-phi} and Proposition~\ref{Computing_nu} to characterize \emph{distributionally robust consideration sets} (DRCS), defined as the set of actions chosen with positive probability at an optimal solution $x^\star$ to (\ref{Max_Lag_phi_DRO_phi}):
	$$
	A(x^\star)
	\triangleq
	\{a\in A:x_a^\star>0\}.
	$$
	Building on the consideration-set result of \cite{Caplinetal2019}, the next proposition extends their analysis to a decision maker who faces ambiguity over the prior.
	\begin{proposition}[Distributionally Robust Consideration Sets (DRCS)]\label{Robust_Consideration_Sets}Let $(x^*,\lambda^*,\eta^*)$ be an optimal solution to problem (\ref{Max_Lag_phi_DRO_phi}) with $\lambda^{\ast}>0$. Assume that $\phi^{\ast}$ is differentiable at $s^\star(\theta)={-H(x^\ast,
			\theta)-\eta^*\over \lambda^*}$ for every $\theta\in\Theta$, and let  $\nu$ be the worst-case probability given  by $\nu(\theta)\triangleq \mu(\theta)\times \phi^{\ast \prime}\left({-H(x^\ast,
			\theta)-\eta^*\over \lambda^*}\right)$ for all $\theta\in \Theta.$  Then the solution $x^*$ is optimally robust if and only if 
		\begin{equation}\label{Consideration_Set_Condition}
			\sum_{\theta \in \Theta} \frac{e^{u(a,\theta)/\xi} \nu(\theta)}{\sum_{b \in A} x^*_b e^{ u(b,\theta)/\xi}} \leq 1\quad \forall a\in A,
		\end{equation}
		with equality if $a\in A(x^*)$.
	\end{proposition}

	The proof combines Theorem~\ref{Thm_DRO-RI-phi} and Proposition~\ref{Computing_nu} with the Lagrangian argument of \cite{Caplinetal2019}, replacing the nominal prior $\mu$ by the worst-case prior $\nu^\star$. The first-order conditions of (\ref{Max_Lag_phi_DRO_phi}) then jointly determine the optimal choice probabilities $x^\star$, the worst-case prior $\nu^\star$, and the associated robust consideration set $A(x^\star)$. Since $\phi^*$ is non-decreasing, $\nu\ge0$ automatically, so no positivity assumption on $\phi^{*\prime}$ is needed: the proposition covers divergences whose worst-case prior does not have full support, such as the Cressie--Read family with $k>1$, for which $\phi_{cr}^{*k}$ is continuously differentiable but $\phi_{cr}^{*k\prime}(s)=0$ for $s\le-1/(k-1)$.	\smallskip

\smallskip
The next corollary specializes Proposition~\ref{Robust_Consideration_Sets} to the Kullback--Leibler case.
	\begin{corollary}\label{Corollary_DRCS}Let $(x^*,\lambda^*,\eta^*)$ be an optimal solution to problem (\ref{Max_Lag_phi_DRO_phi}) with $\lambda^{\ast}>0$. Consider the Kullback-Leibler divergence $\phi_{kl}$. Then the solution $x^*$ is distributionally  robust if and only if 
		\begin{equation}\label{Consideration_Set_Condition2}
			\sum_{\theta \in \Theta} \left(\frac{e^{u(a,\theta)/\xi-H(x^*,\theta)/\lambda} }{\EE_\mu\!\left[e^{-H(x^*,\theta)/\lambda}\right]\sum_{b \in A} x^*_b e^{u(b,\theta)/\xi}}\right)\mu(\theta) \leq 1\quad \forall a\in A,
		\end{equation}
		with equality if $a\in A(x^*)$.
	\end{corollary}
	
	Intuitively, the DM evaluates action $a$ using the adjusted payoff
	$$
	\frac{u(a,\theta)}{\xi}-\frac{H(x^\star,\theta)}{\lambda},$$
	so that inclusion in $A(x^\star)$ depends on utility relative to the state-specific robust continuation value. The resulting double-Shannon specification is closely related to the model studied in \citet{Hansen_et_al_2022}.
\smallskip

As noted in Remark~\ref{remark_AVaR_smoothness}, the AV$@$R-generating divergence introduced in Section~\ref{s3} is non-smooth at the boundary $t=1/\beta$, and its convex conjugate $\phi^{*}$ is not differentiable at $s=0$. The differentiability required by Proposition~\ref{Robust_Consideration_Sets} therefore fails at any state tied at the V$@$R threshold, so that proposition does not directly cover the AV$@$R case. We close this gap by working directly with the reduced Rockafellar--Uryasev formulation in (\ref{concentration_ex1})--(\ref{R_VaR}).
\begin{corollary}[Robust consideration sets under AV$@$R ambiguity]\label{Corollary_AVaR_DRCS}
	Let $\phi$ be the AV$@$R-generating divergence with level $\beta\in(0,1)$, let $x^\star$ solve $\max_{x\in\Delta_n}\{-\mathcal{R}(x)\}$ in (\ref{R_VaR}), and let $\eta^\star$ be an associated optimizer in (\ref{concentration_ex1}), where $\tilde H(x,\theta)\triangleq-H(x,\theta)$. Define
	\begin{equation}\label{AVaR_nu}
		\nu^\star(\theta)
		\triangleq
		\begin{cases}
			\mu(\theta)/\beta,
			& \tilde H(x^\star,\theta)>\eta^\star,\\
			t_\theta\,\mu(\theta)/\beta,
			& \tilde H(x^\star,\theta)=\eta^\star,\\
			0,
			& \tilde H(x^\star,\theta)<\eta^\star,
		\end{cases}
	\end{equation}
	with $t_\theta\in[0,1]$. Then $x^\star$ is optimally robust if and only if there exist $\{t_\theta\}$ such that $\nu^\star$ is a probability measure and
	\begin{equation}\label{Consideration_Set_Condition_AVaR}
		\sum_{\theta\in\Theta}
		\nu^\star(\theta)
		\frac{e^{u(a,\theta)/\xi}}
		{\sum_{b\in A}x_b^\star e^{u(b,\theta)/\xi}}
		\leq 1,
		\qquad
		\forall a\in A,
	\end{equation}
	with equality for every $a\in A(x^\star)$.
\end{corollary}

The weights $t_\theta$ matter only on the threshold set $\Theta_0\triangleq\{\theta:\tilde H(x^\star,\theta)=\eta^\star\}$. Because the AV$@$R functional depends on $\tilde H(x^\star,\cdot)$ only through the separable sum $\sum_\theta\mu(\theta)(\tilde H(x^\star,\theta)-\eta^\star)_+$, the subgradient of $(\cdot)_+$ at a tie can be chosen state by state, so the $t_\theta$ are state-specific in general. When $\Theta_0$ has positive $\mu$-mass they need not be unique, and the worst-case prior may be set-valued; when $\Theta_0=\emptyset$, $\nu^\star$ is unique.

The proof in Appendix~\ref{Proofs} parallels that of Proposition~\ref{Robust_Consideration_Sets}, but works directly with subgradients of the reduced AV$@$R problem in $(x,\eta)$. The important distinction in (\ref{AVaR_nu}) is the discrete-distribution analogue of the standard treatment of mass points at the V$@$R  threshold (\citealp{RockafellarUryasev}). Unlike the Kullback--Leibler worst-case prior in Corollary~\ref{Corollary_DRCS}, which has full support, the AV$@$R worst-case prior in (\ref{AVaR_nu}) assigns positive mass only to states satisfying
\[
\tilde H(x^\star,\theta)\geq\eta^\star,
\]
with potentially fractional mass on states exactly at the threshold. Thus AV$@$R induces a truncation of the state space on the prior-ambiguity side, mirroring the hard attention thresholds generated by Cressie--Read costs on the attention-technology side in Section~\ref{s52}. Section~\ref{s51avar} illustrates this mechanism numerically and documents a cascading ``successive-pooling'' pattern as $\beta$ decreases.	
	
	\subsection{Application: Finding the best alternative}\label{sec:good_alt}We apply Proposition~\ref{Robust_Consideration_Sets} to the problem studied by \cite{Caplinetal2019}, interpreted as the insurance take-up setting of Section~\ref{s1}. A consumer chooses among $N$ coverage plans, $A=\{1,\ldots,N\}$, without knowing which plan best matches her underlying risk exposure (\citealp{BrownJeon2021}). Acquiring enough information to evaluate this match is costly, and, following \cite{AlaryGollierTreich2013}, the consumer also faces Knightian ambiguity about the underlying loss distribution. Exactly one plan is the good option and yields utility $u_G$, while the remaining plans are bad and yield utility $u_B$, with $u_G>u_B$. Setting $\Theta=A$, let $\theta_i$ denote the state in which option $i$ is good. The state-contingent utilities are then given by:
	$$
	u(i, \theta_j) = \begin{cases}
		u_{G}  & i=j, \\
		u_{B}& i\neq j.
	\end{cases}$$

	Here $\mu(\theta_i)$ denotes the prior probability that option $i$ yields the good prize. Without loss of generality, order the states so that
	\[
	\mu_i\equiv\mu(\theta_i)\geq \mu(\theta_{i+1})\equiv\mu_{i+1}.
\]
	The DM acquires information under the Shannon attention cost with parameter $\xi>0$. Following \cite{Caplinetal2019}, we apply the payoff transformation
	
	\begin{equation}\label{Payoffs}
		e^{u(i,\theta_j)/\xi}=\left\{\begin{array}{rr}
			e^{\frac{u_{G}}{\xi} }\equiv \pi(1+\delta) & \text { if } i=j, \\
			e^{\frac{u_{B}}{\xi} }\equiv \pi& \text { if } i \neq j
		\end{array}\right.
	\end{equation}
	with $\pi, \delta>0$. 
	\smallskip
	
	The optimal policy depends on $\delta$ but not on $\pi$. Hence, increases in $u_G-u_B$ and reductions in $\xi$ affect behavior only through their effect on $\delta$. Focusing on the Kullback--Leibler case, Example~\ref{Kullback_Leibler} gives the worst-case probability
	\[
	\nu(\theta_i)=\frac{\mu(\theta_i)e^{-H(x,\theta_i)/\lambda}}{\sum_j\mu(\theta_j)e^{-H(x,\theta_j)/\lambda}}.
	\]
	
	Using
	\[
	H(x,\theta_i)=\xi\log\big[\pi(1+\delta x_i)\big],
\]
	and substituting this expression into the necessary and sufficient conditions of Proposition~\ref{Robust_Consideration_Sets}. Since $\xi$ enters $H(x,\theta_i)$ multiplicatively while $\lambda$ enters the worst-case-prior exponent as $1/\lambda$, it is convenient to work with the dimensionless multiplier $\bar\lambda\triangleq\lambda/\xi$, so that the two combine only through this ratio; with
	\[
	\alpha\triangleq 1+\frac{1}{\bar\lambda}=1+\frac{\xi}{\lambda},
	\]
	yields an explicit threshold characterization. The algebra is deferred to the proof of Proposition~\ref{Consideration_Sets} in Appendix~\ref{Proofs}. Writing $K\triangleq|A(x^*)|$ and
	$$T_\alpha(K)\triangleq\left(\frac{\sum_{i=1}^K\mu(\theta_i)^{1/\alpha}}{K+\delta}\right)^{\alpha}\qquad(\alpha\ge1),$$
	inclusion in the consideration set is governed by comparing $\mu(\theta_K)$ to $T_\alpha(K)$.
	
	\begin{lemma}\label{Lemma_Sbeta} $T_\alpha(K)$ is non-increasing in $\alpha$, for every $K\le N$: if $1\le\alpha\le\alpha'$, then $T_{\alpha'}(K)\le T_\alpha(K)$.
	\end{lemma}

\begin{proposition}\label{Consideration_Sets}
	If $\mu(\theta_N)>T_\alpha(N)$, let $K=N$. Otherwise, let $K<N$ be the unique integer satisfying
	\begin{equation}\label{Inequalities_Consideration}
		\mu(\theta_K)>T_\alpha(K)\geq\mu(\theta_{K+1}).
	\end{equation}
	Then
	\[
	A(x^\star)=\{1,\ldots,K\},
	\]
	and the optimal unconditional choice probabilities satisfy
	\begin{equation}\label{Optimal_policy}
		x_i^\star=\frac{(\mu(\theta_i))^{1/\alpha}(K+\delta)-\sum_{j\in A(x^\star)}(\mu(\theta_j))^{1/\alpha}}{\delta\sum_{j\in A(x^\star)}(\mu(\theta_j))^{1/\alpha}},\qquad	i\in A(x^\star).
	\end{equation}
	Moreover, let $B$ denote the consideration set generated by the benchmark RI problem in which $\mu$ is treated as the correct prior. Then
	\[
	B\subseteq A(x^\star).
	\]
\end{proposition}

Some remarks are in order. First, at $\alpha=1$,
\[
T_1(K)=\frac{\sum_{i\leq K}\mu(\theta_i)}{K+\delta},
\]
so the boundary condition reduces to the familiar criterion
\[
\mu(\theta_N)>\frac{1}{N+\delta}
\]
of \cite{Caplinetal2019}. When $\alpha>1$, Lemma~\ref{Lemma_Sbeta} shows that $T_\alpha(N)$ is weakly smaller, implying that prior robustness weakly expands the consideration set. Thus, the first part of Proposition~\ref{Consideration_Sets} extends Theorem~1 of \cite{Caplinetal2019} by making explicit, through (\ref{Optimal_policy}), how the optimal policy depends on the robustness parameter $\alpha$. As $\bar\lambda\to\infty$ (equivalently $\lambda\to\infty$ for fixed $\xi$), corresponding to $\rho\to0$ and hence $\nu\to\mu$, we have $1/\alpha\to1$, and the DRO-RI solution converges to the benchmark characterization in \cite[Thm.~1]{Caplinetal2019}.	For every finite $\bar\lambda>0$, the solution generally differs from
	the benchmark because the decision maker evaluates policies under a distorted prior. At the opposite extreme, as $\bar\lambda\to0$ and $\rho$ becomes large, the worst-case prior increasingly concentrates on the worst-performing states.
	
Second, the set inclusion $B\subseteq A(x^\star)$ follows from the same single-crossing logic underlying \cite{Caplinetal2019}. In particular, condition~(\ref{Inequalities_Consideration}) identifies $K$ as the largest integer satisfying
	\[
	\mu(\theta_K)>T_\alpha(K),
	\]
	so, for every fixed $\alpha\geq1$, the set of indices satisfying this inequality is an initial segment $\{1,\ldots,K\}$. Let $K_1$ denote the cutoff under the benchmark case $\alpha=1$, and let $K$ denote the cutoff under prior ambiguity. By Lemma~\ref{Lemma_Sbeta}, the case $K<K_1$ is ruled out, implying 
\[
B	\subseteq A(x^\star),
\]
	so prior robustness weakly expands the consideration set.
	\smallskip

Third, from an economic standpoint, $B\subseteq A(x^*)$ reflects that the DRO-RI decision maker is more conservative about \emph{excluding} alternatives: because the worst-case prior $\nu^*$ can reallocate mass toward poorly-served states, she cannot write off a marginal, low-prior alternative as confidently as under the trusted prior $\mu$, so $A(x^*)$ weakly contains more alternatives than the standard approach. This is the consideration-set analogue of \cite{AlaryGollierTreich2013}'s finding that Knightian ambiguity raises demand for insurance and self-insurance: ambiguity about which plan fits leads the consumer to examine more plans, not fewer.
	
	Finally, we note that our result is related to \cite{Hansen_et_al_2022}. In Sections~\ref{s11} and~\ref{s3} we discussed the difference between our constrained formulation and the penalty formulation of \cite{Hansen_et_al_2022}. However, in the present Kullback--Leibler example the two approaches yield the same threshold $T_\alpha(K)$ and the same economic implication: ambiguity with respect to the prior weakly expands the consideration set. Conditional on the relevant shadow-price parameter, both formulations reduce to the same pointwise minimization; the difference lies only in how that parameter is determined. The four broader points of departure from \cite{Hansen_et_al_2022} discussed in Section~\ref{s11} are therefore independent of this particular worked example.	
	\subsubsection{A numerical example}\label{s51num} Next, we show how Proposition \ref{Consideration_Sets} works in a concrete example where $N=5$ represents the number of coverage plans, the prior distribution is
$$
\mu=(\mu_1,\ldots,\mu_5)=(0.42,\,0.23,\,0.16,\,0.11,\,0.08),
$$
and $\delta=1$ (i.e. $(u_G-u_B)/\xi=\ln 2$). For ease of exposition, we normalize $\xi=1$ for this example, so that the multiplier $\bar\lambda=\lambda/\xi$ of Section~\ref{sec:good_alt} coincides numerically with $\lambda$; the column reported below as $\lambda$ should accordingly be read as $\bar\lambda$.\footnote{Recall $\alpha\triangleq1/\bar\lambda+1$: $\alpha=1$ ($\bar\lambda=\infty$, $\rho=0$) recovers the trusted-prior benchmark of \cite{Caplinetal2019}, while $\alpha>1$ corresponds to a positive robustness budget $\rho>0$, with $\bar\lambda=1/(\alpha-1)\downarrow0$ as $\alpha\uparrow\infty$.}

Table \ref{table:numeric_example} reports the cutoff $K^*$ for five different values of $\alpha$ together with the resulting optimal policy $x^*$ from (\ref{Optimal_policy}).

{\small
	\begin{table}[ht]
		\caption{Robust consideration sets as ambiguity increases. Entries left blank for $i\notin A(x^*)$ with $\mu=(.42,.23,.16,.11,.08)$, $\delta=1$}
		\centering
		\begin{tabular}{c c c c c c c c c}
			\hline\hline
			$\alpha$ & $\bar\lambda = 1/(\alpha-1)$ & $K^*$ & $A(x^*)$ & $x_1^*$ & $x_2^*$ & $x_3^*$ & $x_4^*$ & $x_5^*$ \\[0.5ex]
			\hline
			$1.0$ & $\infty$ & $2$ & $\{1,2\}$ & $0.938$ & $0.062$ & -- & -- & -- \\
			$1.5$ & $2.00$ & $2$ & $\{1,2\}$ & $0.797$ & $0.203$ & -- & -- & -- \\
			$2.0$ & $1.00$ & $3$ & $\{1,2,3\}$ & $0.697$ & $0.256$ & $0.047$ & -- & -- \\
			$3.0$ & $0.50$ & $4$ & $\{1,2,3,4\}$ & $0.571$ & $0.285$ & $0.139$ & $0.005$ & -- \\
			$5.0$ & $0.25$ & $5$ & $\{1,\ldots,5\}$ & $0.431$ & $0.268$ & $0.180$ & $0.094$ & $0.027$ \\
			[1ex]
			\hline
		\end{tabular}
		\label{table:numeric_example}
	\end{table}
}

At $\alpha=1$ the trusting consumer studies only plans $\{1,2\}$, placing $93.8\%$ of attention on plan $1$. As ambiguity increases, $A(x^*)$ expands as Proposition \ref{Consideration_Sets} predicts --- plan $3$ enters at $\alpha=2$, plan $4$ at $\alpha=3$, the full menu at $\alpha=5$ --- confirming $B\subseteq A(x^*)$ with strict inclusion for sufficiently large $\alpha$. The table also shows a complementary effect: conditional on inclusion, weight on the most-likely plan falls monotonically, from $0.938$ to $0.431$, as attention flattens across the newly-included plans. Thus, prior ambiguity induces broader attention across the menu.

\smallskip

\begin{figure}[ht]
	\centering
	\includegraphics[width=0.95\linewidth]{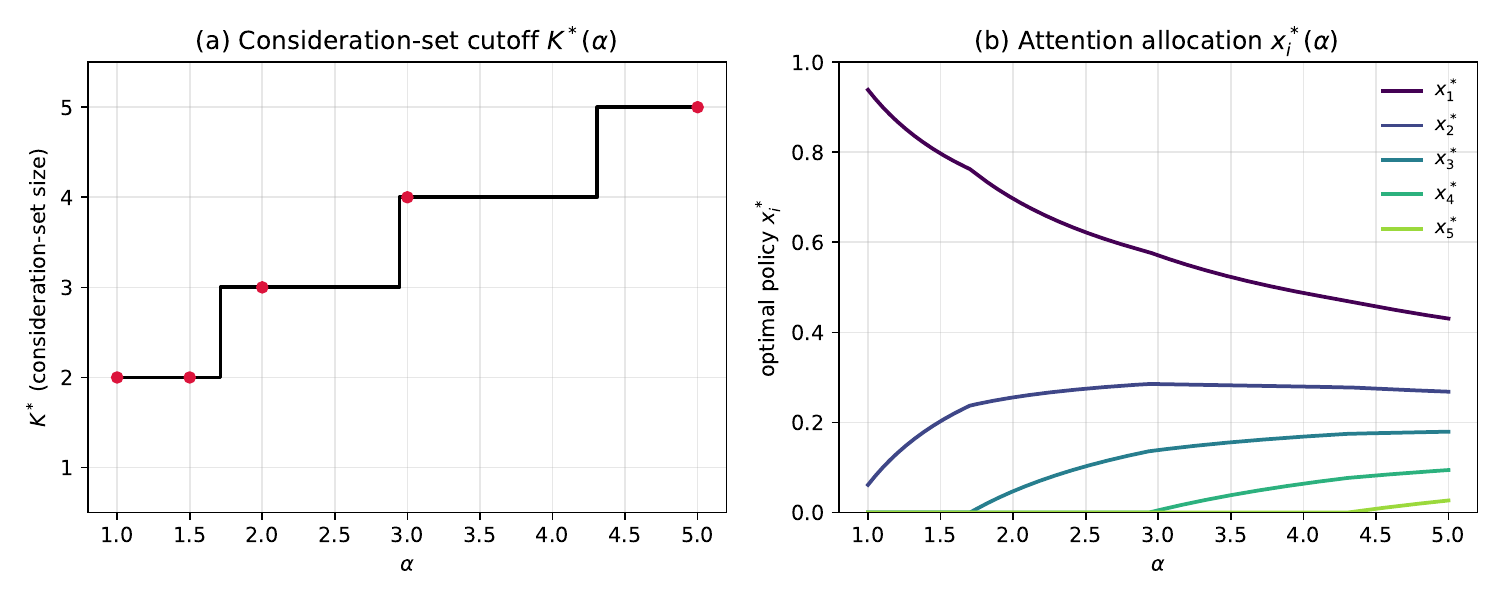}
	\caption{Robust consideration sets in the insurance take-up example. (a) Consideration-set cutoff $K^*$ as a function of $\alpha$. (b) Optimal policy $x_i^*$ for each of the $N=5$ plans as a function of $\alpha$. Red dots in (a) mark the five values of $\alpha$ reported in Table \ref{table:numeric_example}.}
	\label{fig:consideration_sets}
\end{figure}

Figure \ref{fig:consideration_sets} plots both effects continuously in $\alpha$: panel (a) shows the cutoff $K^*(\alpha)$ as the step function of Proposition \ref{Consideration_Sets}; panel (b) shows $x_i^*(\alpha)$, with newly-admitted plans entering continuously at zero while $x_1^*(\alpha)$ declines throughout.

\subsection{Application: AV$@$R-robust attention and successive pooling}\label{s51avar}

We now  revisit the insurance take-up example under the AV$@$R uncertainty set of Section~\ref{s3}, with confidence level $\beta\in(0,1]$, replacing the Kullback--Leibler ambiguity considered above. To allow a direct comparison with the Kullback--Leibler example, we keep the parametrization
\[
N=5,\qquad
\mu=(0.42,0.23,0.16,0.11,0.08),\qquad
\delta=1,
\]
fixed, isolating the effect of changing the shape of the uncertainty set. As with $\alpha$ in Section~\ref{s51num}, $\beta$ indexes robustness, but in the opposite direction: smaller $\beta$ corresponds to greater robustness, while $\beta=1$ recovers the trusted-prior benchmark.

\begin{figure}[ht]
	\centering
	\includegraphics[width=0.85\textwidth]{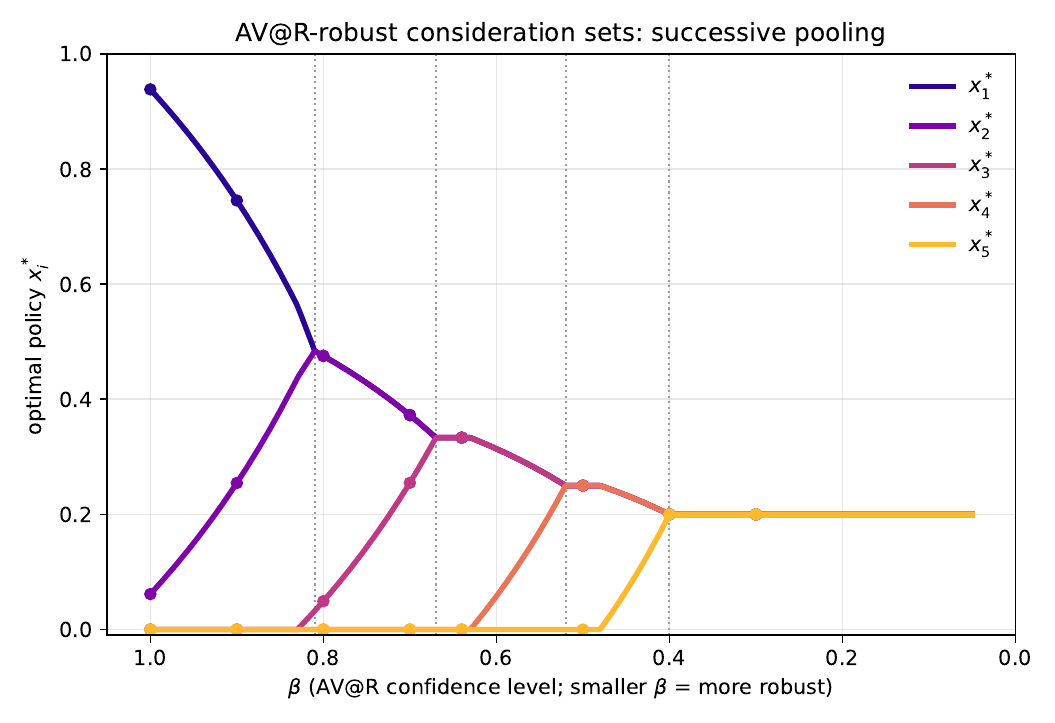}
	\caption{AV$@$R-robust attention allocation $x_i^*$ by confidence level in the insurance take-up example. Robustness increases rightward as the confidence level decreases. Dots indicate the eight values in Table \ref{table:avar_numeric}; dotted vertical lines mark the pooling thresholds $\beta_2,\ldots,\beta_5$ of Remark \ref{remark_pooling}. As the confidence level falls, plans progressively pool, ultimately reaching the uniform policy in Proposition \ref{Prop_AVaR_uniform}.}
	\label{fig:avar_pooling}
\end{figure}

Table~\ref{table:avar_numeric} reports the optimal policy $x^\star(\beta)$, computed numerically from (\ref{R_VaR}), for a decreasing sequence of AV$@$R confidence levels $\beta$. Figure \ref{fig:avar_pooling} plots the full path; two features stand out, neither present in the Kullback--Leibler case of \S \ref{s51num}.

{\small
	\begin{table}[ht]
		\caption{AV$@$R-robust consideration sets and successive pooling ($\mu=(.42,.23,.16,.11,.08)$, $\delta=1$)}
		\centering
		\begin{tabular}{c c c c c c c}
			\hline\hline
			$\beta$ & $x_1^*$ & $x_2^*$ & $x_3^*$ & $x_4^*$ & $x_5^*$ & pooled group \\[0.5ex]
			\hline
			$1.00$ & $0.938$ & $0.062$ & $0$     & $0$     & $0$     & none ($K=2$, as in Table \ref{table:numeric_example}) \\
			$0.90$ & $0.745$ & $0.255$ & $0$     & $0$     & $0$     & none \\
			$0.80$ & $0.475$ & $0.475$ & $0.049$ & $0$     & $0$     & $\{1,2\}$ \\
			$0.70$ & $0.373$ & $0.373$ & $0.255$ & $0$     & $0$     & $\{1,2\}$ \\
			$0.64$ & $0.333$ & $0.333$ & $0.333$ & $0$     & $0$     & $\{1,2,3\}$ \\
			$0.50$ & $0.250$ & $0.250$ & $0.250$ & $0.250$ & $0$     & $\{1,2,3,4\}$ \\
			$0.40$ & $0.200$ & $0.200$ & $0.200$ & $0.200$ & $0.200$ & $\{1,2,3,4,5\}$ \\
			$0.30$ & $0.200$ & $0.200$ & $0.200$ & $0.200$ & $0.200$ & $\{1,2,3,4,5\}$ \\
			[1ex]
			\hline
		\end{tabular}
		\label{table:avar_numeric}
	\end{table}
}

Two features of the AV$@$R solution are worth emphasizing. First, rather than merely flattening as $\beta$ falls, as under Kullback--Leibler ambiguity, the optimal policy exhibits \emph{successive pooling}. The unconditional choice probabilities merge in stages: plans $1$ and $2$ at $\beta=0.81$, plan $3$ joins them at $\beta=0.67$, plan $4$ at $\beta=0.52$, and plan $5$ at $\beta=0.40$, at which point the policy is fully pooled. These pooling levels are given in closed form in Remark~\ref{remark_pooling} below. This is the AV$@$R analogue of the hard attention thresholds studied in Section~\ref{s52}: here, it is the difference $x_i^\star-x_j^\star$, rather than an individual choice probability, that is driven exactly to zero. Economically, these pooling events reflect the tail-truncated worst-case prior characterized in Corollary~\ref{Corollary_AVaR_DRCS}. As $\beta$ decreases, the adversary places positive weight on an expanding set of adverse states, inducing the DM to equalize choice probabilities across an increasing number of plans.

Second, this ``cascade effect'' has a sharp analytical endpoint.

\begin{proposition}[Extreme ambiguity aversion and uniform attention]\label{Prop_AVaR_uniform}
	Fix $x\in\Delta_n$. As $\beta\downarrow0$,
	\[\mathcal{R}(x)\uparrow\max_{\theta\in\Theta}\tilde H(x,\theta)=-\min_{\theta\in\Theta}H(x,\theta).
	\]
	Hence,
	\[\max_{x\in\Delta_n}{-\mathcal{R}(x)}\longrightarrow
	\max_{x\in\Delta_n}\min_{\theta\in\Theta}H(x,\theta)
	\]
	as $\beta\downarrow0$. Moreover, the limit is attained at a finite confidence level: for every $\beta\le\min_{\theta\in\Theta}\mu(\theta)$, $\mathcal{R}(x)=\max_{\theta\in\Theta}\tilde H(x,\theta)$ for all $x\in\Delta_n$, and hence $\max_{x\in\Delta_n}\{-\mathcal{R}(x)\}=\max_{x\in\Delta_n}\min_{\theta\in\Theta}H(x,\theta)$.
\end{proposition}

The proof in Appendix~\ref{Proofs} shows that the classical convergence of AV$@$R to the essential supremum (\citealp{RockafellarUryasev,Shapiro_et_al_2013}) is reached at a finite confidence level: once $\beta\le\min_\theta\mu(\theta)$, the AV$@$R ambiguity set (\ref{Uncertainty_CVaR}) contains the point mass on the worst state. In specific applications, the optimal policy can reach its limit well before this bound, as the next remark shows for the insurance example. Proposition~\ref{Consideration_Sets} yields the same limiting allocation in the Kullback--Leibler case: its optimal policy in (\ref{Optimal_policy}) satisfies $x_i^\star\to1/N$ as $\alpha\to\infty$. Thus, the divergence specifications considered here converge under extreme ambiguity to the same prior-independent uniform policy. What differs is the path to that limit: smooth tilting under Kullback--Leibler ambiguity, hard thresholds under Cressie--Read divergence, and state-truncated successive pooling under AV$@$R.

To gain some intuition about this result, consider  the insurance take-up example,
\[
H(x,\theta_j)
=
\xi\log\!\bigl[\pi(1+\delta x_j)\bigr].
\]
Since $\xi>0$ and $\xi\log\pi$ is common across states, maximizing
$\min_j H(x,\theta_j)$ is equivalent to maximizing
\[
\min_j \log(1+\delta x_j),
\]
which depends on $x$ only through $x_j$ and is strictly increasing with the same functional form for every $j$. Therefore, maximizing the minimum statewise indirect utility is equivalent to maximizing $\min_j x_j$ over the simplex. The unique solution is
\[x^\star=\left(\frac1N,\ldots,\frac1N\right),\]
independently of the prior $\mu$.

\begin{remark}[Pooling thresholds in the insurance example]\label{remark_pooling}
	The pooling levels in Table~\ref{table:avar_numeric} can be computed exactly from Corollary~\ref{Corollary_AVaR_DRCS}. Suppose plans $1,\ldots,k$ are pooled, $x_1^\star=\cdots=x_k^\star>0$, while $x_j^\star=0$ for $j>k$. Then states $\theta_1,\ldots,\theta_k$ share the same loss $\tilde H(x^\star,\theta_i)$, while the states $\theta_j$, $j>k$, have strictly larger loss and therefore receive the maximal weight $\mu(\theta_j)/\beta$. Since $e^{u(i,\theta)/\xi}/\sum_bx_b^\star e^{u(b,\theta)/\xi}$ differs across included plans only through the term $(1+\delta)/(1+\delta x_i^\star)$ in state $\theta_i$, the equalities in (\ref{Consideration_Set_Condition_AVaR}) hold for all $i\le k$ only if $\nu^\star(\theta_1)=\cdots=\nu^\star(\theta_k)=\big(1-\sum_{j>k}\mu(\theta_j)/\beta\big)/k$. This common weight must respect the cap $\nu^\star(\theta_i)\le\mu(\theta_i)/\beta$, which binds first at $i=k$. Hence plans $1,\ldots,k$ can be pooled only if
	\[
	\beta\;\le\;\beta_k\;\triangleq\;k\,\mu(\theta_k)+\sum_{j>k}\mu(\theta_j).
	\]
	Since $\beta_k-\beta_{k+1}=k\big(\mu(\theta_k)-\mu(\theta_{k+1})\big)\ge0$, the thresholds are ordered, which produces the cascade. For $\mu=(.42,.23,.16,.11,.08)$ this gives $\beta_2=0.81$, $\beta_3=0.67$, $\beta_4=0.52$, and $\beta_5=0.40$, matching the pooling points in Table~\ref{table:avar_numeric} and Figure~\ref{fig:avar_pooling}. With $k=N$, every state is tied at the uniform policy and the condition is also sufficient (take $t_\theta=\beta/(N\mu(\theta))\in[0,1]$): the uniform policy is optimal if and only if $\beta\le N\mu(\theta_N)$, which equals $0.40$ here, well above the bound $\min_\theta\mu(\theta)=0.08$ guaranteed by Proposition~\ref{Prop_AVaR_uniform} for general payoffs.
\end{remark}

\subsection{Generalized attention costs: the Cressie--Read case}\label{s52}

We return to the composite model of Section~\ref{s34} and examine consideration sets when the attention technology departs from Shannon entropy. We focus on the Cressie--Read family,
\[\phi_1=\phi_{cr}^k,\qquad k>1,\]
using the same divergence family introduced earlier for prior ambiguity, but now at the \emph{attention-cost} layer.
	\subsubsection{A power-law generalization of the RI logit}
	
	Using
	\[\phi_{cr}^{k\prime}(t)=\frac{t^{k-1}-1}{k-1},\]
	the first-order condition for (\ref{Generalized_H}) yields a closed-form primal solution, providing a natural counterpart to the dual representation in Lemma~\ref{Generalized_H_Dual}.
	
	\begin{lemma}\label{CR_attention_rule}
		Let $\phi_1=\phi_{cr}^k$ with $k>1$. For every $x\in\Delta_n$ and $\theta\in\Theta$, the optimal conditional choice rule solving (\ref{Generalized_H}) is
		\begin{equation}\label{CR_logit}
			p_a(\theta)=	x_a\left[1+(k-1)\frac{u(a,\theta)-\eta_\theta}{\xi}
			\right]_{+}^{\frac{1}{k-1}},\qquad	\forall a\in A,
		\end{equation}
		where $\eta_\theta\in\RR$ is the unique Lagrange multiplier associated with the constraint:
				\[\sum_{a\in A}p_a(\theta)=1.\]
		Moreover, as $k\downarrow1$,
	\[\left[1+(k-1)\frac{u(a,\theta)-\eta_\theta}{\xi}\right]^{\frac{1}{k-1}}
		\longrightarrow
		\exp\left(
		\frac{u(a,\theta)-\eta_\theta}{\xi}
		\right),
		\]
		so (\ref{CR_logit}) converges to the Shannon logit in (\ref{RIlogitprobs}).
	\end{lemma}
The proof in Appendix~\ref{Proofs} solves the Lagrangian first-order condition directly for $p_a(\theta)$, rather than working through the conjugate $\phi_1^*$ as in Lemma~\ref{Generalized_H_Dual}. The two characterizations are equivalent by Fenchel's equality.

\smallskip

Expression~(\ref{CR_logit}) generalizes the RI logit to a \emph{power-law} attention rule, tilting conditional choice probabilities by a power rather than an exponential function of $u(a,\theta)-\eta_\theta$. Unlike the Shannon case, whose tilting factor is strictly positive, the Cressie--Read factor is exactly zero whenever
\[u(a,\theta)\leq\eta_\theta-\frac{\xi}{k-1}.
\]
Thus, Cressie--Read attention costs generate a \emph{hard} attention threshold even before introducing any outer $\phi_2$-robustness layer.	
	\subsubsection{Consideration sets under Cressie--Read attention costs}
	
	We apply (\ref{CR_logit}) to the ``good alternative'' example of Section~\ref{sec:good_alt}. Define
\[z_j\triangleq\left[1+(k-1)\frac{u_B-\eta_j}{\xi}\right]_{+}^{1/(k-1)}\]
	and
\[w_j\triangleq\left[1+(k-1)\frac{u_G-\eta_j}{\xi}\right]_{+}^{1/(k-1)},\]
	which represent the attention intensities assigned, respectively, to an incorrect and the correct choice in state $\theta_j$. The normalization condition becomes
	\begin{equation}\label{CR_normalization}
		x_jw_j+(1-x_j)z_j=1.
	\end{equation}
	
	Unlike the Kullback--Leibler case, in which
	\[\frac{w_j}{z_j}=e^{(u_G-u_B)/\xi}\]
	is constant across states, for $k\neq1$ the ratio depends on the state-specific multiplier $\eta_j$. Hence (\ref{CR_normalization}) is generally implicit, and the closed-form characterization in Proposition~\ref{Consideration_Sets} is a special feature of the Shannon exponential structure.
	
	Nevertheless, (\ref{CR_logit}) yields a sharp statewise attention threshold. Conditional on $x_i>0$, option $i$ receives positive conditional choice probability in state $\theta_j$ if and only if
	\begin{equation}\label{CR_threshold}
		u(i,\theta_j)>\eta_j-\frac{\xi}{k-1}.
	\end{equation}

	Thus, Cressie--Read costs introduce an additional extensive-margin mechanism: even an option belonging to the unconditional consideration set can receive zero conditional probability in particular states. The cutoff in (\ref{CR_threshold}) is governed by the curvature parameter $k$, providing a power-law counterpart to the Shannon consideration-set structure studied by  \cite{Caplinetal2019}.
	\begin{example}[Quadratic attention costs, $k=2$]\label{Quadratic_Example}
		For $k=2$,
		\[\phi_1=\phi_{cr}^2(t)=\frac{1}{2}(t-1)^2,
		\]
		the modified $\chi^2$-divergence reported in Table~\ref{table:examples}. In this case, (\ref{CR_logit}) becomes
\[p_j(\theta_j)=x_j\left[
		1+\frac{u(j,\theta_j)-\eta_j}{\xi}
		\right]_+.\]
		
Suppose first that
	\[\xi>x_j(u_G-u_B),\]
so that all terms inside the positive-part operator are strictly positive. Then the normalization condition (\ref{CR_normalization}) yields
\[\eta_j=x_j u_G+(1-x_j)u_B,\]
		and hence
		\begin{eqnarray}
			p_j(\theta_j)
			&=&x_j\left[1+(1-x_j)\frac{u_G-u_B}{\xi}\right]_{+},	\label{Quadratic_pjj}\\
			p_i(\theta_j)&=&x_i\left[
			1-x_j\frac{u_G-u_B}{\xi}\right]_{+},\qquad\forall i\neq j.\label{Quadratic_pij}
		\end{eqnarray}
		
At the threshold
\[\xi=x_j(u_G-u_B),\]
		the off-diagonal probabilities reach zero. For
\[\xi\leq x_j(u_G-u_B),\]
the optimal conditional choice rule is
\[p_j(\theta_j)=1,\qquad p_i(\theta_j)=0 \quad\forall i\neq j.
		\]
		Thus, quadratic Cressie--Read attention costs generate an exact statewise attention threshold: sufficiently low information costs lead the decision maker to identify the good option with probability one. This contrasts with the Shannon specification, under which conditional choice probabilities remain strictly positive.
	\end{example}
	\smallskip
	
Combining Cressie--Read attention costs with a nondegenerate outer layer of prior ambiguity proceeds as in Proposition~\ref{Robust_Consideration_Sets}: the nominal prior $\mu$ is replaced by the worst-case prior $\nu^\star$, while the conditional choice rule continues to satisfy (\ref{CR_logit})--(\ref{CR_threshold}). The optimal reference distribution $x^\star$ and worst-case prior $\nu^\star$ are jointly characterized by Theorem~\ref{Composite_Thm}.

We next consider the quadratic attention cost of Example~\ref{Quadratic_Example}, corresponding to $k=2$, combined with a Kullback--Leibler outer ambiguity layer, $\phi_2=\phi_{kl}$. This specification remains tractable in closed form. The corresponding model with an AV$@$R outer ambiguity layer does not generally admit a closed-form solution and can instead be solved numerically.
\subsubsection{A fully worked composite example}\label{s523}

We return to the good-alternative example with $\phi_1=\phi_{cr}^2$, as in Example~\ref{Quadratic_Example}, and now set $\phi_2=\phi_{kl}$ in the composite problem (\ref{Composite_DRO-RI})--(\ref{Composite_Dual}). Define
\[\gamma\triangleq\frac{u_G-u_B}{\xi},\]
the analogue of the Shannon-case parameter $\delta$ in Section~\ref{s51num}.

\smallskip
\noindent\emph{Closed form for $H_{\phi_1}$.} Solving the statewise inner problem in Lemma~\ref{Generalized_H_Dual} for $\phi_1=\phi_{cr}^2$ (derivation in Appendix \ref{Proofs}) gives, for each $j$,
\begin{equation}\label{H_phi1_closed}
	H_{\phi_1}(x,\theta_j)=\begin{cases}
		u_B+(u_G-u_B)\,x_j+\dfrac{\xi\gamma^2}{2}\,x_j(1-x_j), & x_j\leq 1/\gamma,\\[1.5ex]
		u_G-\dfrac{\xi}{2}\cdot\dfrac{1-x_j}{x_j}, & x_j> 1/\gamma.
	\end{cases}
\end{equation}

The first branch corresponds to the interior, smooth regime of Example~\ref{Quadratic_Example}, in which $p_i(\theta_j)>0$ for every $i$. The second branch corresponds to the hard-threshold regime characterized in (\ref{CR_threshold}). Once $x_j$ exceeds $1/\gamma$, the DM allocates sufficient attention to alternative $j$ in state $\theta_j$ that
\[p_i(\theta_j)=0\qquad\forall i\neq j.
\]
In this branch $\partial H_{\phi_1}(x,\theta_j)/\partial x_j=\xi/(2x_j^2)$: once the conditional choice rule has become deterministic, further changes in $x_j$ affect $H_{\phi_1}(x,\theta_j)$ only through the divergence cost of concentrating the reference distribution on that deterministic rule, not through any further change in $p_j(\theta_j)$ itself. The two branches meet smoothly at $x_j=1/\gamma$: both their values and first derivatives coincide at the threshold.
\smallskip

\noindent\emph{Outer layer in closed form.}
Because $\phi_2=\phi_{kl}$ has conjugate $
\phi_2^*(s)=e^s-1,$ the maximization over $\eta$ in (\ref{Composite_Dual}) can be solved explicitly; see Appendix~\ref{Proofs}. The composite problem therefore reduces to the KL-robust certainty equivalent of the generalized indirect utility $H_{\phi_1}$:
\begin{equation}\label{Composite_KL_Entropic}
	\upsilon^\star_{\phi_1,\phi_{kl}}(\zeta)=\max_{x\in\Delta_n}\left\{-\zeta\log\EE_\mu\!\left[e^{-H_{\phi_1}(x,\theta)/\zeta}\right]\right\},
\end{equation}
with associated worst-case prior, by the same Gibbs (exponential-tilting) argument used throughout \S \ref{s3}--\S \ref{s34},
\begin{equation}\label{Composite_KL_nu}
	\nu^*(\theta_j)=\frac{\mu(\theta_j)\,e^{-H_{\phi_1}(x^*,\theta_j)/\zeta}}{\sum_{l}\mu(\theta_l)\,e^{-H_{\phi_1}(x^*,\theta_l)/\zeta}}.
\end{equation}
Expression~(\ref{Composite_KL_Entropic}) is concave in $x$ by Theorem~\ref{Composite_Thm}, and $\zeta$ plays the same role as the multiplier $\lambda$ in the base model. As $\zeta\to\infty$, deviations from the reference prior become prohibitively costly, so that $\nu^\star\to\mu$ and the model converges to the non-robust specification of Example~\ref{Quadratic_Example}. Smaller values of $\zeta$ penalize deviations from $\mu$ less heavily and therefore correspond to greater robustness to prior misspecification.
\subsubsection{Composite numerical illustration}\label{Composite_Numeric_Example} Let $N=3$ alternatives, $\mu=(0.5,0.3,0.2)$, $u_B=0$, $u_G=2$, $\xi=1$, so that $\gamma=2$ and the hard-threshold cutoff of (\ref{H_phi1_closed}) is $x_j>1/2$. Table \ref{table:composite_numeric} reports the solution $x^*$ of (\ref{Composite_KL_Entropic}), obtained numerically for a decreasing sequence of the penalty parameter $\zeta$, corresponding to increasing robustness to prior misspecification.

{\small
	\begin{table}[ht]
		\caption{Composite CR-attention/KL-prior model: optimal policy and hard-threshold regime as outer ambiguity increases ($\mu=(.5,.3,.2)$, $u_B=0$, $u_G=2$, $\xi=1$, $\gamma=2$)}
		\centering
		\begin{tabular}{c c c c c c}
			\hline\hline
			$\zeta$ & $x_1^*$ & $x_2^*$ & $x_3^*$ & state-1 regime & $\nu^*$ \\[0.5ex]
			\hline
			$\infty$ & $0.575$ & $0.370$ & $0.055$ & hard threshold ($p_1(\theta_1)\!=\!1$) & $(0.500,0.300,0.200)$ \\
			$5$      & $0.536$ & $0.337$ & $0.127$ & hard threshold ($p_1(\theta_1)\!=\!1$) & $(0.464,0.305,0.231)$ \\
			$2.5$    & $0.510$ & $0.325$ & $0.165$ & hard threshold ($p_1(\theta_1)\!=\!1$) & $(0.437,0.311,0.252)$ \\
			$2$      & $0.499$ & $0.322$ & $0.179$ & interior (just below cutoff) & $(0.426,0.315,0.260)$ \\
			$1$      & $0.448$ & $0.323$ & $0.229$ & interior & $(0.395,0.322,0.283)$ \\
			$0.5$    & $0.404$ & $0.327$ & $0.269$ & interior & $(0.370,0.328,0.302)$ \\
			[1ex]
			\hline
		\end{tabular}
		\label{table:composite_numeric}
	\end{table}
}

As $\zeta\to\infty$, the worst-case prior converges to the reference prior, $\nu^\star\to\mu$, and the solution converges to that of Example~\ref{Quadratic_Example}. In particular,
\[
x_1^\star=0.575>\frac{1}{\gamma}=\frac{1}{2},
\]
so state $\theta_1$ lies in the hard-threshold branch of (\ref{H_phi1_closed}), with
\[
p_i(\theta_1)=0
\qquad
\forall i\neq1.
\]
Such exact zeros in conditional choice probabilities cannot arise under Shannon attention costs.

As $\zeta$ decreases, two effects emerge. First, attention shifts toward lower-prior alternatives: $x_3^\star$ rises from $0.055$ to $0.269$, echoing the diversification mechanism in Proposition~\ref{Consideration_Sets}. Second, and specific to the composite model, $x_1^\star$ falls back through the hard-threshold cutoff $1/\gamma=1/2$ at approximately $\zeta^\star=2.03$. At this point, the conditional attention rule in state $\theta_1$ switches from the deterministic, zero-mistake branch to the interior branch of (\ref{H_phi1_closed}). Greater robustness to prior misspecification makes concentrating attention on state $\theta_1$ less attractive as the worst-case prior shifts probability toward states $\theta_2$ and $\theta_3$.

\begin{figure}[ht]
	\centering
	\includegraphics[width=0.85\textwidth]{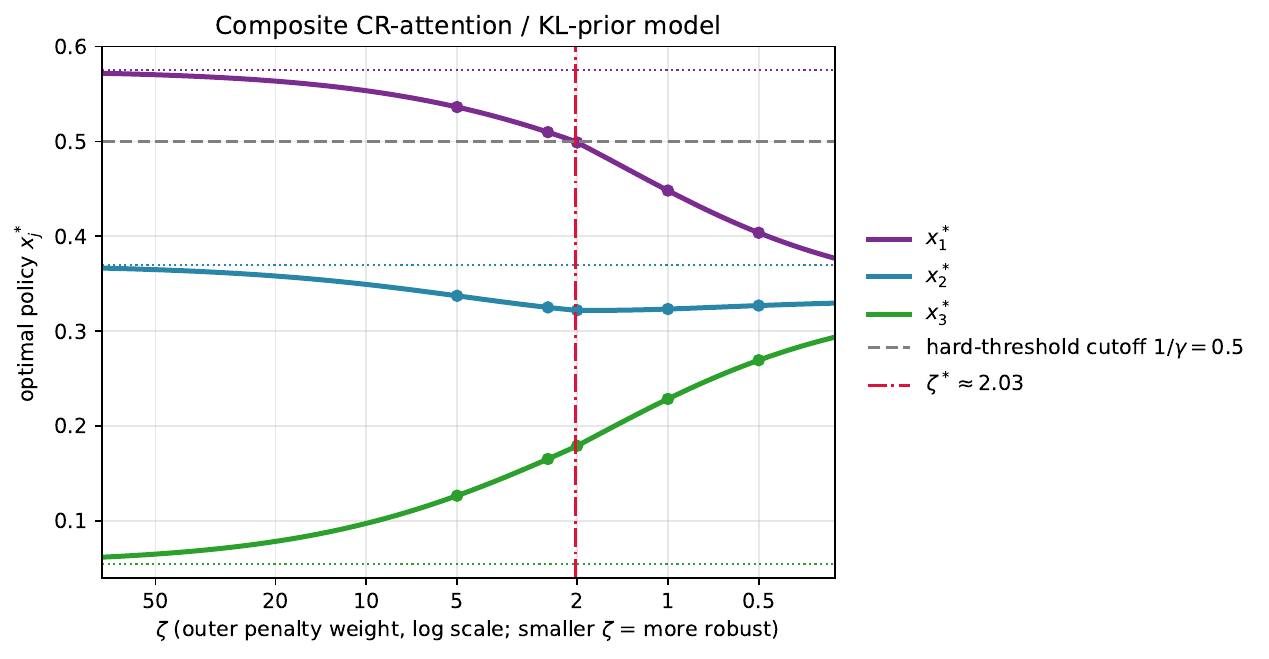}
	\caption{Composite Cressie--Read attention / KL-robust prior model: optimal policy $x_j^*$ as a function of the outer penalty weight $\zeta$ (log scale, decreasing left to right, so that robustness increases rightward). The dashed line is the hard-threshold cutoff $1/\gamma$; the dash-dotted line marks the crossover $\zeta^*$ at which $x_1^*$ re-enters the interior (smooth) regime. Dotted lines mark the $\zeta\to\infty$ benchmark of Example \ref{Quadratic_Example}.}
	\label{fig:composite_zeta}
\end{figure}

This threshold crossing is a genuinely composite phenomenon: the Cressie--Read attention-cost layer generates the hard threshold, while the outer prior-ambiguity layer can make it optimal for the decision maker to retreat from that threshold.
Figure \ref{fig:composite_zeta} plots $x_1^*,x_2^*,x_3^*$ against $\zeta$, making the regime switch explicit: the $x_1^*$ path crosses the hard-threshold cutoff $1/\gamma=0.5$ (dashed) at $\zeta^*\approx2.03$ (dash-dotted). Dotted lines mark the $\zeta\to\infty$ benchmark of Example \ref{Quadratic_Example}.

\section{Final Remarks}\label{s6}
In this paper, we study RI when the decision maker faces ambiguity about the prior. We introduce a distributionally robust rational inattention (DRO-RI) model and show that, for a general $\phi$-divergence ball around the benchmark prior, the robust problem reduces to a finite-dimensional concave program that jointly determines the attention policy and the worst-case prior. We exploit this structure to derive necessary and sufficient conditions for robust consideration sets and show that the DRO-RI problem admits a natural interpretation through robust risk measures, linking prior ambiguity to tail-sensitive decision criteria.

The choice of divergence shapes behavior, not only its degree. In the insurance take-up example, Kullback--Leibler ambiguity expands the consideration set and flattens attention smoothly; AV$@$R ambiguity truncates the state space and produces successive pooling of choice probabilities at explicit thresholds; and Cressie--Read attention costs generate exact zeros in conditional choice. All three converge to the same prior-independent uniform policy under extreme ambiguity. The composite and nested extensions show that these mechanisms can be combined --- separating robustness in the attention technology from robustness in beliefs, or treating the baseline prior itself as ambiguous --- without sacrificing tractability.

Several questions remain open. Our consideration-set results concern the support of the optimal policy, not how much information is acquired; how prior ambiguity affects mutual information, and whether constraint and multiplier formulations respond differently to the cost of attention, are natural next steps. A second direction is empirical: because robustness distorts the revealed default rule relative to observed choice frequencies in a menu-dependent way, stochastic choice data with controlled priors may separate prior ambiguity from a fixed misspecified prior. Finally, extending the analysis to dynamic rational inattention raises the question of how updating under ambiguity interacts with costly learning. We leave these questions for future work.
	\bibliographystyle{plainnat}
	\bibliography{References_DRO_RI_misspecification_October}
\appendix
	\section{For online publication}\label{Proofs}
	The following lemma is key in proving  the equivalence Proposition \ref{Robust_max_min_equiv}.
	\begin{lemma} [Donsker-Varadhan Lemma]
		\begin{equation}
			\log \EE_p e^{f(z)}=\max _{q \in \Delta(Z)}\left\{\EE_q f(z)-D(q \| p)\right\}
		\end{equation}	
		where $f$ is a real value function on $Z$ and $D(q\|p)=\sum_{z}q(z)\log(q(z)/p(z))$ is the relative entropy (Kullback--Leibler divergence) of $q$ with respect to $p$, the maximum being taken over $q\ll p$.
	\end{lemma}
	\subsection{Proof of Proposition \ref{Robust_max_min_equiv}} Write
	$$F(P,\nu)\triangleq\sum_{\theta\in\Theta}\Big[\sum_{a\in A}p_a(\theta)u(a,\theta)\Big]\nu(\theta)-\xi C(P,\nu)$$
	for the objective of (\ref{Program1_robust_max_min}). The proof proceeds in three steps.

	\emph{Step 1 (pointwise equality).} For each state $\theta \in \Theta$ and any strategy $x\in\Delta_n$, applying the Donsker-Varadhan Lemma with $p=x$ and $f(a)=u_a(\theta)/\xi$, and multiplying both sides by $\xi$, gives
	$$H(x,\theta)=\xi\log\sum_{a\in A}x_ae^{u_a(\theta)/\xi}=\max _{p_a(\theta) \in \Delta(A)}\left\{\sum_{a\in A}u_a(\theta)p_a(\theta)-\xi D(p(\theta) \| x)\right\}$$
	where $p(\theta)=(p_a(\theta))_{a\in A}$, consistent with $H(x,\theta)\triangleq\xi\log(x^\top e^{u(\theta)/\xi})$ and the coefficient $\xi$ (not $1/\xi$) on $C(P,\mu)$ in (\ref{Mutual_information_cost}). Fix $\nu\in\mathcal B$. Taking an expectation over $\theta$ with respect to $\nu$ and then maximizing over $x \in\Delta_n$, the left side becomes $\max_{x\in \Delta_{n}}\EE_\nu(H(x,\theta))$, while
	\begin{eqnarray*}
		\max_{x\in\Delta_n}\EE_{\nu}(H(x,\theta))&=&\max_{x\in\Delta_n,\,P\in\mathcal{P}}\sum_{\theta\in \Theta}\left(\sum_{a\in A}u_a(\theta)p_a(\theta)\right)\nu(\theta)-\xi\EE_{\nu}(D(p(\theta)\|x))\\
		&=&\max_{P\in\mathcal{P}}\sum_{\theta\in \Theta}\left(\sum_{a\in A}u_a(\theta)p_a(\theta)\right)\nu(\theta)-\xi C(P,\nu)\\
		&=&\max_{P\in\mathcal P}F(P,\nu),
	\end{eqnarray*}
	where the second equality uses that, for fixed $P$, $-\xi\EE_\nu[D(p(\theta)\|x)]$ is maximized over $x\in\Delta_n$ at $x=\sum_\theta p(\theta)\nu(\theta)$ (the marginal of $p(\theta)$ induced by $\nu$), at which point $\xi\EE_\nu[D(p(\theta)\|x)]=\xi C(P,\nu)$ by definition (\ref{Mutual_information_cost}), consistent with Proposition~\ref{MM_result}. This establishes the pointwise equality
	\begin{equation}
		\max_{P\in\mathcal P}F(P,\nu)=\max_{x\in\Delta_n}\EE_\nu[H(x,\theta)]\qquad\text{for every }\nu\in\mathcal B. \tag{$\star$}
	\end{equation}

	\emph{Step 2 (first application of Sion's theorem).} $\mathcal P$ is compact and convex (a product of simplices), $\mathcal B$ is compact and convex by hypothesis, $F(P,\nu)$ is concave in $P$ for fixed $\nu$ (its first term is linear in $P$, and $C(P,\nu)$, being mutual information, is convex in the channel $P$), and convex in $\nu$ for fixed $P$ (its first term is linear in $\nu$, and $C(P,\nu)$ is concave in the input distribution $\nu$); $F$ is also continuous on $\mathcal P\times\mathcal B$. Sion's (\citeyear{Sion_1958}) minimax theorem therefore gives
	$$\max_{P\in\mathcal P}\min_{\nu\in\mathcal B}F(P,\nu)=\min_{\nu\in\mathcal B}\max_{P\in\mathcal P}F(P,\nu).$$
	Substituting ($\star$) into the right-hand side,
	$$\min_{\nu\in\mathcal B}\max_{P\in\mathcal P}F(P,\nu)=\min_{\nu\in\mathcal B}\max_{x\in\Delta_n}\EE_\nu[H(x,\theta)].$$

	\emph{Step 3 (second application of Sion's theorem).} $\Delta_n$ is compact and convex, $\mathcal B$ is compact and convex, $\EE_\nu[H(x,\theta)]$ is concave in $x$ (since $H(x,\theta)=\xi\log(x^\top e^{u(\theta)/\xi})$ is the composition of the concave function $\log(\cdot)$ with an affine function of $x$, and expectation preserves concavity), and linear in $\nu$ for fixed $x$. A second application of Sion's theorem gives
	$$\min_{\nu\in\mathcal B}\max_{x\in\Delta_n}\EE_\nu[H(x,\theta)]=\max_{x\in\Delta_n}\min_{\nu\in\mathcal B}\EE_\nu[H(x,\theta)].$$

	Chaining Steps 2 and 3,
	$$\max_{P\in\mathcal P}\min_{\nu\in\mathcal B}F(P,\nu)=\max_{x\in\Delta_n}\min_{\nu\in\mathcal B}\EE_\nu[H(x,\theta)],$$
	which is the claimed equivalence (\ref{Program1_robust_max_min})--(\ref{Program1_robust_max_min_2}).\eproof
	
	\subsection{Proof of Theorem \ref{Thm_DRO-RI-phi}} We adapt the Lagrangian duality arguments in \cite[Cor. 3]{BenTaletal_2013}. We first study the inner minimization program. Using the fact that $$\min_{\nu\in \mathcal{B}_\phi(\mu)}\EE_{\nu}(H(x,\theta))=-\max_{\nu\in \mathcal{B}_\phi(\mu)}-\EE_{\nu}(H(x,\theta)).$$
	The associated Lagrangian for $\max_{\nu\in \mathcal{B}_\phi(\mu)}-\EE_{\nu}(H(x,\theta))$ is
	\begin{equation}\label{Lagrangian_phi}
		\mathcal{L}(\nu,\lambda, \eta)=-\EE_{\nu}(H(x,\theta))+\lambda\left(\rho-\sum_{\theta\in \Theta}\mu(\theta)\phi\left(\nu(\theta)\over \mu(\theta)\right)\right)+\eta\left(1-\sum_{\theta\in \Theta}\nu(\theta)\right).
	\end{equation}
	The dual objective function is
	\begin{equation}\label{Dual_lagrangian}
		g(\lambda,\eta)=\max_{\nu\geq 0}\mathcal{L}(\nu,\lambda,\eta).
	\end{equation}
	\begin{eqnarray}
		g(\lambda, \eta) &=&\lambda\rho+\eta+\max_{\nu\geq 0}\sum_{\theta\in \Theta}\left(-\nu(\theta)H(x,\theta)-\lambda\mu(\theta)\phi\left(\nu(\theta)\over \mu(\theta)\right)-\eta\nu(\theta)\right)\nonumber,\\
		&=&\lambda\rho+\eta+\sum_{\theta\in \Theta}\max_{\nu(\theta)\geq 0}\left(-\nu(\theta)H(x,\theta)-\lambda\mu(\theta)\phi\left(\nu(\theta)\over \mu(\theta)\right)-\eta\nu(\theta)\right) \nonumber,\\
		&=& \lambda\rho+\eta+\sum_{\theta\in \Theta}\max_{\nu(\theta)\geq 0}\left(\nu(\theta)(-H(x,\theta)-\eta)-\lambda\mu(\theta)\phi\left(\nu(\theta)\over \mu(\theta)\right)\right) \nonumber,\nonumber\\
		&=& \lambda\rho+\eta+\sum_{\theta\in \Theta}\mu(\theta)\max_{t\geq 0}\left(t(-H(x,\theta)-\eta)-\lambda\phi\left(t\right)\right) \nonumber\\
		&=& \lambda\rho+\eta+\sum_{\theta\in \Theta}\mu(\theta)(\lambda\phi)^*\left(-H(x,\theta)-\eta\right) \label{Dual_Langrangian_Equality}
	\end{eqnarray}
	where in the  equality (\ref{Dual_Langrangian_Equality}) we define $t\triangleq{\nu(\theta)\over \mu(\theta)}$ for all $\theta\in \Theta.$ Using the definition of $\phi^*$ and noting that $(\lambda\phi)^*(s)=\lambda\phi^*(s/\lambda)$ for $\lambda\geq 0$, where we define $0\phi^*(s/0)\triangleq (0\phi)^*(s)$, which equals $0$ if $s\leq 0$ and $+\infty$ if $s>0.$ 
	Since $\rho>0$, $\mu$ has full support, and $\phi(1)=0$, the point $\nu=\mu$ satisfies $I_\phi(\mu,\mu)=0<\rho$ and lies in the relative interior of $\Delta(\Theta)$, so Slater's condition holds for the convex program $\max_{\nu\in\mathcal B_\phi(\mu)}-\EE_\nu(H(x,\theta))$. Hence there is no duality gap and the dual infimum is attained (\citealp[Thm.~28.2]{Rockafellar1970}): $\max_{\nu\in\mathcal B_\phi(\mu)}-\EE_\nu(H(x,\theta))=\min_{\lambda\geq0,\eta}g(\lambda,\eta)$, and therefore $\min_{\nu\in\mathcal B_\phi(\mu)}\EE_\nu(H(x,\theta))=\max_{\lambda\geq0,\eta}\{-g(\lambda,\eta)\}$ for every $x\in\Delta_n$.
	Plugging in equality (\ref{Dual_Langrangian_Equality}) in (\ref{DR0-RI-phi}) we obtain 
	{\footnotesize
		\begin{eqnarray}\label{Eq_duality}
			\max_{x\in \Delta_n}\min_{\nu\in \mathcal{B}_\phi(\mu)}\{\EE_{\nu}(H(x,\theta))\}&=&\max _{x \in \Delta_n, \lambda \geq 0, \eta}\left\{-\eta-\rho \lambda-\lambda \sum_{\theta\in \Theta} \mu(\theta) \phi^{*}\left(\frac{-H(x, \theta)-\eta}{\lambda}\right)\right\}.
	\end{eqnarray}}
	For concavity, note that $(s,\lambda)\mapsto\lambda\phi^*(s/\lambda)$ is jointly convex on $\RR\times\RR_+$ (it is the perspective of the convex function $\phi^*$) and non-decreasing in $s$, because $\operatorname{dom}\phi\subseteq[0,\infty)$ implies that $\phi^*$ is non-decreasing. Since $H(x,\theta)$ is concave in $x$, $-H(x,\theta)-\eta$ is jointly convex in $(x,\eta)$, so each term $\lambda\phi^*\big((-H(x,\theta)-\eta)/\lambda\big)$ is jointly convex in $(x,\lambda,\eta)$. Hence the objective in (\ref{Eq_duality}) is jointly concave in $(x,\lambda,\eta)$.
	\eproof
	\subsection{Proof of Proposition \ref{Computing_nu}} By the proof of Theorem \ref{Thm_DRO-RI-phi}, strong duality holds and the dual optimum $(\lambda^\star,\eta^\star)$ is attained, so the KKT conditions are necessary and sufficient for $\nu^\star$ to be a worst-case prior at $x^\star$. Stationarity of the Lagrangian (\ref{Lagrangian_phi}) in each $\nu(\theta)$ requires $t^\star\triangleq\nu^\star(\theta)/\mu(\theta)$ to maximize $t\,(-H(x^\star,\theta)-\eta^\star)-\lambda^\star\phi(t)$ over $t\geq0$; for $\lambda^\star>0$ this is equivalent to $s^\star(\theta)\in\partial\phi(t^\star)$, i.e.\ to (\ref{RP_1}) by Fenchel's equality. Complementary slackness, $\lambda^\star\big(\rho-I_\phi(\nu^\star,\mu)\big)=0$ with $\lambda^\star>0$, gives (\ref{RP_2}), and primal feasibility gives (\ref{RP_3}).\eproof

	\subsection{Proof of Corollary \ref{optimized_value}} Under Assumption~\ref{Duality_Regularity}, and since $\phi^\ast$ is differentiable at $s^\star(\theta)$ for every $\theta\in\Theta$, the optimality condition with respect to $\lambda$ (cf.\ the derivation of (\ref{FOC2}) in the proof of Proposition~\ref{Robust_Consideration_Sets}, which is the same computation) is
	$$\rho+\EE_{\mu}(\phi^\ast(s(\theta)))+\frac1\lambda\EE_{\mu}\left(\phi^{\ast\prime}(s(\theta))\big(H(x,\theta)+\eta\big)\right)=0.$$
	Evaluating at $x^\star, \lambda^\star$, and $\eta^\star$, multiplying through by $-\lambda^\star$, and using $\EE_\mu(\phi^{\ast\prime}(s^\star(\theta)))=\sum_{\theta\in\Theta}\nu^\star(\theta)=1$ (Proposition~\ref{Computing_nu}), we find:
	\begin{equation}\label{optimized_equivalence_algebra}
		-\lambda^\star\rho-\eta^\star-\lambda^\star\EE_{\mu}(\phi^\ast(s^\star(\theta)))=\EE_{\mu}(\phi^{\ast\prime}(s^\star(\theta))H(x^\star,\theta)).
	\end{equation}
	Noting that $\EE_{\mu}(\phi^{\ast\prime}(s^\star(\theta))H(x^\star,\theta))=\EE_{\nu^\star}(H(x^\star,\theta))$, and that the LHS of (\ref{optimized_equivalence_algebra}) equals $\upsilon^\star$ by (\ref{Max_Lag_phi_DRO_phi}), we conclude $\upsilon^\star=\EE_{\nu^\star}(H(x^\star,\theta))$.\eproof
	
	\subsection{Proof of Lemma \ref{Generalized_H_Dual}} Fix $\theta\in\Theta$. By definition,
	$$H_{\phi_1}(x,\theta)=\max_{p(\theta)\in \Delta_n}\left\{\sum_{a\in A}p_a(\theta)u(a,\theta)-\xi\sum_{a\in A}x_a\phi_1\left(\frac{p_a(\theta)}{x_a}\right)\right\}.$$
	Relaxing the simplex constraint on $p(\theta)$ to $p_a(\theta)\geq 0$ for all $a$ (recall $\phi_1(t)=+\infty$ for $t<0$ by convention) and dualizing $\sum_{a\in A}p_a(\theta)=1$ with multiplier $\eta_\theta\in\RR$, the Lagrangian is
	$$\mathcal{L}(p(\theta),\eta_\theta)=\sum_{a\in A}p_a(\theta)u(a,\theta)-\xi\sum_{a\in A}x_a\phi_1\left(\frac{p_a(\theta)}{x_a}\right)+\eta_\theta\left(1-\sum_{a\in A}p_a(\theta)\right).$$
	Since $\phi_1$ is convex, the objective is concave in $p(\theta)$, and the remaining constraint is linear and satisfied by $p(\theta)=x$, which lies in the relative interior of the domain of the objective restricted to $\operatorname{supp}(x)$. (Coordinates with $x_a=0$ are handled by the convention of Definition \ref{Phi_Divergence}: when $\phi_1$ is superlinear, $\lim_{t\to\infty}\phi_1(t)/t=+\infty$, as for $\phi_{kl}$ and $\phi_{cr}^k$ with $k>1$, they are forced to $p_a(\theta)=0$ and contribute $0=x_a\phi_1^*(\cdot)$ to both the primal and the dual.) Hence strong duality obtains. Setting $t\triangleq p_a(\theta)/x_a$, the pointwise maximization over $p_a(\theta)\geq 0$ for each $a\in A$ gives
	$$\max_{t\geq 0}\left\{x_at\big(u(a,\theta)-\eta_\theta\big)-\xi x_a\phi_1(t)\right\}=\xi x_a\,\phi_1^{*}\!\left(\frac{u(a,\theta)-\eta_\theta}{\xi}\right).$$
	Summing over $a\in A$, adding $\eta_\theta$, and minimizing over $\eta_\theta$ yields (\ref{Generalized_H_Dual_Eq}). For $\phi_1=\phi_{kl}$ we have $\phi_{kl}^{*}(s)=e^{s}-1$; substituting and solving the first-order condition in $\eta_\theta$ explicitly (as in the derivation of (\ref{KL_Expression})) recovers $H_{\phi_1}(x,\theta)=\xi\log\sum_{a\in A}x_ae^{u(a,\theta)/\xi}$.\eproof

	\subsection{Proof of Lemma \ref{CR_attention_rule}} Fix $\theta\in\Theta$ and $\phi_1=\phi_{cr}^k$ with $k>1$. As in the proof of Lemma \ref{Generalized_H_Dual}, relax the simplex constraint on $p(\theta)$ to $p_a(\theta)\geq0$ and dualize $\sum_{a\in A}p_a(\theta)=1$ with multiplier $\eta_\theta\in\RR$; the objective $\sum_ap_a(\theta)u(a,\theta)-\xi\sum_ax_a\phi_1(p_a(\theta)/x_a)$ is concave in $p(\theta)$, and, as in that proof (restricting attention to $\operatorname{supp}(x)$ if $x$ has zero coordinates), strong duality obtains and the primal optimum is characterized by the pointwise first-order conditions. For each $a\in A$ with $p_a(\theta)>0$, differentiating the Lagrangian with respect to $p_a(\theta)$ and using $\phi_{cr}^{k\prime}(t)=(t^{k-1}-1)/(k-1)$ gives
	$$u(a,\theta)-\xi\,\phi_1'\!\left(\frac{p_a(\theta)}{x_a}\right)-\eta_\theta=0
	\quad\Longleftrightarrow\quad
	\xi\,\frac{(p_a(\theta)/x_a)^{k-1}-1}{k-1}=u(a,\theta)-\eta_\theta,$$
	so that
	$$\left(\frac{p_a(\theta)}{x_a}\right)^{k-1}=1+(k-1)\frac{u(a,\theta)-\eta_\theta}{\xi}.$$
	Since $k>1$, the map $t\mapsto t^{k-1}$ is a strictly increasing bijection from $[0,\infty)$ onto $[0,\infty)$, so solving for $p_a(\theta)/x_a$ gives $\left[1+(k-1)(u(a,\theta)-\eta_\theta)/\xi\right]^{1/(k-1)}$ whenever the bracketed term is non-negative. When it is instead negative, the unconstrained stationary point falls outside the domain $p_a(\theta)\geq0$ (recall $\phi_{cr}^k(t)=+\infty$ for $t<0$), so the constrained optimum is the boundary solution $p_a(\theta)=0$. Combining both cases via the positive-part operator yields (\ref{CR_logit}).

	The left-hand side of the normalization $\sum_{a\in A}p_a(\theta)=1$, viewed as a function of $\eta_\theta$ through (\ref{CR_logit}), is continuous and (weakly, and strictly wherever the corresponding term is positive) decreasing in $\eta_\theta$, tends to $\sum_ax_a\cdot\infty=\infty$ as $\eta_\theta\to-\infty$ and to $0$ as $\eta_\theta\to+\infty$; hence there is a unique $\eta_\theta\in\RR$ solving $\sum_ap_a(\theta)=1$.

	Finally, for the limit $k\downarrow1$, write $z\triangleq(u(a,\theta)-\eta_\theta)/\xi$ and $\epsilon\triangleq k-1\downarrow0$. Then $\left[1+\epsilon z\right]^{1/\epsilon}=\exp\!\left(\frac{\log(1+\epsilon z)}{\epsilon}\right)\to\exp(z)$ as $\epsilon\to0$, since $\log(1+\epsilon z)/\epsilon\to z$ by L'H\^opital's rule (or the standard first-order expansion of $\log$ near $1$). This is exactly the Shannon tilting factor in (\ref{RIlogitprobs}). The convergence is locally uniform in $z$, and the multiplier $\eta_\theta=\eta_\theta(k)$ is the unique root of the continuous normalization equation above, whose limit as $k\downarrow1$, $\sum_ax_ae^{(u(a,\theta)-\eta)/\xi}=1$, is strictly decreasing in $\eta$ with unique root $\xi\log\sum_ax_ae^{u(a,\theta)/\xi}$. Hence $\eta_\theta(k)$ converges to the Shannon multiplier, and (\ref{CR_logit}) converges to the Shannon logit (\ref{RIlogitprobs}).\eproof

	\subsection{Proof of Theorem \ref{Composite_Thm}} By Lemma \ref{Generalized_H_Dual}, $H_{\phi_1}(x,\theta)$ is a well-defined real-valued function of $(x,\theta)$, concave in $x$ (an infimum of functions affine in $x$). Fix $x\in\Delta_n$ and treat $H_{\phi_1}(x,\theta)$ as a given state function. The inner problem in (\ref{Composite_DRO-RI}) is
	$$\min_{\nu\in\Delta(\Theta)}\left\{\sum_{\theta\in\Theta}\nu(\theta)H_{\phi_1}(x,\theta)+\zeta\sum_{\theta\in\Theta}\mu(\theta)\phi_2\!\left(\frac{\nu(\theta)}{\mu(\theta)}\right)\right\}.$$
	Dualizing the single constraint $\sum_\theta\nu(\theta)=1$ with multiplier $\eta\in\RR$ (the non-negativity of $\nu$ is handled pointwise, as in the proof of Theorem \ref{Thm_DRO-RI-phi}), the Lagrangian is
	$$\mathcal{L}(\nu,\eta)=\sum_\theta\nu(\theta)H_{\phi_1}(x,\theta)+\zeta\sum_\theta\mu(\theta)\phi_2\!\left(\frac{\nu(\theta)}{\mu(\theta)}\right)+\eta\left(1-\sum_\theta\nu(\theta)\right).$$
	Since $\phi_2$ is convex, the objective is convex in $\nu$ and Slater's condition holds ($\nu=\mu$ is interior), so strong duality obtains. Setting $t\triangleq\nu(\theta)/\mu(\theta)$, the pointwise minimization over $\nu(\theta)\geq 0$ for each $\theta$ gives
	$$\min_{t\geq0}\left\{\mu(\theta)\,t\big(H_{\phi_1}(x,\theta)-\eta\big)+\zeta\mu(\theta)\phi_2(t)\right\}=-\zeta\mu(\theta)\,\phi_2^{*}\!\left(\frac{\eta-H_{\phi_1}(x,\theta)}{\zeta}\right).$$
	Summing over $\theta\in\Theta$ and adding $\eta$ gives
	$$\min_{\nu\in\Delta(\Theta)}\{\cdots\}=\max_{\eta\in\RR}\left\{\eta-\zeta\EE_\mu\left[\phi_2^{*}\left(\frac{\eta-H_{\phi_1}(x,\theta)}{\zeta}\right)\right]\right\};$$
	taking $\max_x$ of both sides yields (\ref{Composite_Dual}).
	
	This argument treats $H_{\phi_1}(x,\theta)$ as a fixed, already-concave function of $x$, supplied separately by Lemma \ref{Generalized_H_Dual}, rather than substituting its dual representation $H_{\phi_1}(x,\theta)=\inf_{\eta_\theta}B_\theta(x,\eta_\theta)$, where $B_\theta$ denotes the objective in (\ref{Generalized_H_Dual_Eq}). Such a substitution would not flatten the problem: since $\phi_2^{*}$ is non-decreasing and continuous, $-\zeta\phi_2^{*}\big((\eta-\inf_{\eta_\theta}B_\theta)/\zeta\big)=\inf_{\eta_\theta}\big\{-\zeta\phi_2^{*}\big((\eta-B_\theta)/\zeta\big)\big\}$, so the objective in (\ref{Composite_Dual}) becomes $\max_{x,\eta}\{\eta+\EE_\mu[\inf_{\eta_\theta}(\cdots)]\}$, a max--inf that cannot be merged into a single joint maximization over $(x,\eta,\{\eta_\theta\})$ without an additional minimax argument. We therefore state Theorem \ref{Composite_Thm} as the nested program (\ref{Composite_Dual}); since each inner problem is a one-dimensional convex program, it is no less tractable.
	
	Concavity of (\ref{Composite_Dual}) in $(x,\eta)$ follows because $H_{\phi_1}(x,\theta)$ is concave in $x$ (Lemma \ref{Generalized_H_Dual}), so $\eta-H_{\phi_1}(x,\theta)$ is convex in $(x,\eta)$ jointly (affine in $\eta$, convex in $x$); since $\phi_2^{*}$ is convex and non-decreasing, $\phi_2^{*}((\eta-H_{\phi_1}(x,\theta))/\zeta)$ is convex in $(x,\eta)$ (a non-decreasing convex function of a convex function is convex), and $\eta$ minus a convex function, scaled by $\zeta>0$, is concave. The joint recovery of $(x^*,p^*(\theta),\nu^*)$ follows from the first-order conditions, applied state-by-state to (\ref{Generalized_H_Dual_Eq}) and to the outer layer above, exactly as in Proposition \ref{Computing_nu}.\eproof
	
	See \S \ref{s34} for the reason we do not invoke \cite[Thm.~1]{Jinetal2024} verbatim: their theorem dualizes a single global normalization with two scalars $(\theta_1,\theta_2)$, whereas our conditional attention technology requires $|\Theta|$ separate per-state normalizations, which is exactly the source of the nesting (rather than flattening) just discussed.

	\subsection{Derivation of the closed-form composite CR/KL example (\S\ref{s523})} Solving the state-wise inner problem of Lemma \ref{Generalized_H_Dual} explicitly for $\phi_1=\phi_{cr}^2$ (so $\phi_1^*(s)=s+s^2/2$ for $s\geq-1$, and $\phi_1^*(s)=-1/2$ otherwise) gives, for each $j$ and using $H_{\phi_1}(x,\theta_j)$'s dependence on $x$ only through $x_j$, the two-branch expression (\ref{H_phi1_closed}); the two branches, and their first derivatives, agree at $x_j=1/\gamma$. Because $\phi_2=\phi_{kl}$ has conjugate $\phi_2^*(s)=e^s-1$, the outer maximization over $\eta$ in (\ref{Composite_Dual}) can be solved in closed form exactly as in (\ref{concentration_ex22}), giving (\ref{Composite_KL_Entropic}) with associated worst-case prior (\ref{Composite_KL_nu}) by the same Gibbs (exponential-tilting) argument used throughout \S \ref{s3}--\S \ref{s34}.

	\subsection{Proof of Theorem \ref{Nested_Thm}} Dualizing $\sum_\theta\mu(\theta)=1$ with multiplier $\eta_1$ and $\sum_\theta\hat\mu(\theta)=1$ with multiplier $\eta_2$, and handling non-negativity pointwise, the Lagrangian is
	\begin{align*}
	\mathcal{L}=&\sum_\theta\Big\{\mu(\theta)H(x,\theta)+\psi\mu(\theta)\log\frac{\mu(\theta)}{\hat\mu(\theta)}+\zeta\mu_0(\theta)\phi_2\Big(\frac{\hat\mu(\theta)}{\mu_0(\theta)}\Big)\Big\}\\
	&+\eta_1\Big(1-\sum_\theta\mu(\theta)\Big)+\eta_2\Big(1-\sum_\theta\hat\mu(\theta)\Big).
	\end{align*}
	The objective is jointly convex in $(\mu,\hat\mu)$ (relative entropy is jointly convex and $\phi_2$ is convex), and Slater's condition holds at $(\mu,\hat\mu)=(\mu_0,\mu_0)$, which lies in the relative interior of $\Delta(\Theta)\times\Delta(\Theta)$ since $\mu_0$ has full support; hence there is no duality gap. Since there is no coupling across states, the minimization decomposes into $|\Theta|$ separate two-variable problems, one per $\theta$. Fix $\theta$ and write $(m,h)\triangleq(\mu(\theta),\hat\mu(\theta))$. First minimize over $m\geq 0$ for fixed $h>0$: the first-order condition $H(x,\theta)+\psi[\log(m/h)+1]-\eta_1=0$ gives $m^*(h)=h\,e^{(\eta_1-H(x,\theta))/\psi-1}$, and substituting back,
	$$\min_{m\geq0}\Big\{mH(x,\theta)+\psi m\log\tfrac{m}{h}-\eta_1 m\Big\}=-\psi h\,e^{(\eta_1-H(x,\theta))/\psi-1}.$$
	(This is the same Gibbs variational step as in Example \ref{Kullback_Leibler} and in \citealp{Hansen_et_al_2022}'s eq.~(9), with $h$ in place of their baseline.) Next minimize over $h\geq0$: writing $c\triangleq-\psi e^{(\eta_1-H(x,\theta))/\psi-1}-\eta_2$ for the coefficient of $h$ once the $-\eta_2h$ term (from the second Lagrange term) is included, and using the standard conjugate identity $\min_{h\geq0}\{ch+\zeta\mu_0(\theta)\phi_2(h/\mu_0(\theta))\}=-\zeta\mu_0(\theta)\phi_2^*(-c/\zeta)$, we obtain, for each $\theta$,
	$$\min_{m,h\geq0}\{\cdots\}=-\zeta\mu_0(\theta)\,\phi_2^{*}\!\left(\frac{\eta_2+\psi e^{(\eta_1-H(x,\theta))/\psi-1}}{\zeta}\right).$$
	Summing over $\theta\in\Theta$, adding $\eta_1+\eta_2$, and reparametrizing via $\eta_1\mapsto\eta_1+\psi$ to absorb the $e^{-1}$ term --- i.e., substituting the old dual variable by $\eta_1+\psi$, where $\eta_1$ now denotes the new variable, so that $\psi e^{((\eta_1+\psi)-H)/\psi-1}=\psi e^{(\eta_1-H)/\psi}$, removing the $-1$ in the exponent, while the linear term becomes $(\eta_1+\psi)+\eta_2=\eta_1+\eta_2+\psi$ --- gives
	$$\min_{\hat\mu\in\Delta(\Theta)}\min_{\mu\in\Delta(\Theta)}\{\cdots\}=\max_{\eta_1,\eta_2\in\RR}\left\{\eta_1+\eta_2+\psi-\zeta\EE_{\mu_0}\!\left[\phi_2^{*}\!\left(\frac{\eta_2+\psi e^{(\eta_1-H(x,\theta))/\psi}}{\zeta}\right)\right]\right\},$$
	and taking $\max_x$ of both sides yields (\ref{Nested_Dual}).
	
	For the two limits we work with the primal. By the Gibbs step above, $\min_{\mu\in\Delta(\Theta)}\{\EE_\mu H(x,\theta)+\psi I_{\phi_{kl}}(\mu,\hat\mu)\}=-\psi\log\EE_{\hat\mu}[e^{-H(x,\theta)/\psi}]$, and $J(\mu_0;\psi,\zeta)$ is the maximum over $x\in\Delta_n$ of the value of the inner minimization in (\ref{Nested_Problem}). We use the following elementary fact. Let $K$ be compact, let $a:\Delta_n\times K\to\RR$ be continuous, let $b:K\to[0,\infty]$ and $D:K\to[0,\infty]$ be lower semicontinuous, and let $K_0\triangleq D^{-1}(0)$ be such that $v_\infty(x)\triangleq\min_{k\in K_0}\{a(x,k)+b(k)\}$ is finite. Then $v_t(x)\triangleq\min_{k\in K}\{a(x,k)+b(k)+tD(k)\}$ is non-decreasing in $t$ and $v_t\uparrow v_\infty$ uniformly on $\Delta_n$ as $t\to\infty$, so that $\max_{x\in\Delta_n}v_t(x)\to\max_{x\in\Delta_n}v_\infty(x)$. Indeed, $v_t\le v_\infty$; if $k_t$ minimizes, then $tD(k_t)\le v_\infty(x)-\min_{\Delta_n\times K}a$, so $D(k_t)\to0$ and, by lower semicontinuity of $D$, every limit point $\bar k$ of $(k_t)$ lies in $K_0$; along a subsequence attaining $\liminf_tv_t(x)$ and converging to $\bar k$, lower semicontinuity of $b$ gives $\liminf_tv_t(x)\ge a(x,\bar k)+b(\bar k)\ge v_\infty(x)$. Finally, $|v_t(x)-v_t(x')|\le\max_{k\in K}|a(x,k)-a(x',k)|$, and likewise for $v_\infty$, so all these functions are continuous on the compact set $\Delta_n$ by uniform continuity of $a$, and Dini's theorem upgrades monotone pointwise convergence to uniform convergence.

	For part (i), apply this fact with $t=\zeta$, $K=\Delta(\Theta)$, $a(x,\hat\mu)=-\psi\log\EE_{\hat\mu}[e^{-H(x,\theta)/\psi}]$, $b=0$, and $D=I_{\phi_2}(\cdot,\mu_0)$. Since $\phi_2$ is convex with $\phi_2(1)=\phi_2'(1)=0$ and $\phi_2''(1)>0$, we have $\phi_2(t)>0$ for $t\neq1$, and since $\mu_0$ has full support, $K_0=\{\mu_0\}$. Hence $J(\mu_0;\psi,\zeta)\to\max_x\{-\psi\log\EE_{\mu_0}[e^{-H(x,\theta)/\psi}]\}$, which is \citeauthor{Hansen_et_al_2022}'s eq.~(11) with $\mu_0$ in place of their $\hat\mu$; the argument also shows that the worst-case $\hat\mu^\star\to\mu_0$, while for finite $\zeta$ the outer layer continues to allow $\hat\mu$ to deviate from $\mu_0$.

	For part (ii), apply the same fact with $t=\psi$, $K=\Delta(\Theta)\times\Delta(\Theta)\ni k=(\mu,\hat\mu)$, $a(x,k)=\EE_\mu H(x,\theta)$, $b(k)=\zeta I_{\phi_2}(\hat\mu,\mu_0)$, and $D(k)=I_{\phi_{kl}}(\mu,\hat\mu)$, so that $K_0=\{\mu=\hat\mu\}$ and $v_\infty(x)=\min_{\hat\mu\in\Delta(\Theta)}\{\EE_{\hat\mu}H(x,\theta)+\zeta I_{\phi_2}(\hat\mu,\mu_0)\}$. By the proof of Theorem \ref{Composite_Thm} with $H_{\phi_1}=H$ and nominal prior $\mu_0$, this equals $\max_\eta\{\eta-\zeta\EE_{\mu_0}[\phi_2^*((\eta-H(x,\theta))/\zeta)]\}$, and maximizing over $x$ gives the stated limit; the argument also shows that $\mu^\star-\hat\mu^\star\to0$.\eproof
	
	The following lemma is useful in proving Proposition \ref{Simplified_Dual1}.
	\begin{lemma}\label{dual_CR} For the Cressie-Read family the following holds:
		\begin{equation}
			\phi^{*k}_{cr}(s)=\frac{1}{k}((k-1) s+1)_{+}^{k^{*}}-\frac{1}{k}
		\end{equation}
	\end{lemma}
	\proof By definition,
	\[
	\phi_{cr}^{*k}(s)
	=
	\sup_{t\geq 0}
	\left\{
	st-\phi_{cr}^k(t)
	\right\}.
	\]
	For \(t>0\),
	\[
	\frac{\partial}{\partial t}
	\left[
	st-\phi_{cr}^k(t)
	\right]
	=
	s-\frac{t^{k-1}-1}{k-1}.
	\]
	If
	\[
	s\leq -\frac{1}{k-1},
	\]
	the derivative is non-positive for every \(t\geq0\), so the supremum is attained
	at \(t=0\). If instead
	\[
	s>-\frac{1}{k-1},
	\]
	the first-order condition gives
	\[
	t^\star
	=
	\left((k-1)s+1\right)^{1/(k-1)}.
	\]
	Substituting \(t^\star\) into the objective yields
	\[
	\phi_{cr}^{*k}(s)
	=
	\frac{1}{k}
	\left(
	\left((k-1)s+1\right)_+^{k^\ast}
	-1
	\right),
	\qquad
	k^\ast=\frac{k}{k-1}.
	\]
	\eproof
	\subsection{Proof of Proposition \ref{Simplified_Dual1}} In proving this proposition we follow the proof of Lemma 1 in  \cite{DuchiNamkoong2021} and Lemma \ref{dual_CR} above.  From Theorem \ref{Thm_DRO-RI-phi} we know the following:
	
	\begin{eqnarray*}
		\inf _{\lambda \geq 0, \eta\in \RR}\left\{\lambda \mathbb{E}_{\mu}\left( \phi_{cr}^{*k}\left(\frac{\tilde{H}(x,\theta)-\eta}{\lambda}\right)\right)+\lambda \rho+\eta\right\}&=& \\
		\inf _{\lambda \geq 0, \eta\in \RR}\left\{\frac{(k-1)^{k_{*}}}{k} \lambda^{1-k_{*}} \mathbb{E}_{\mu}\left(\tilde{H}(x,\theta)-\eta+\frac{\lambda}{k-1}\right)_{+}^{k_{*}}+\lambda\left(\rho-\frac{1}{k}\right)+\eta\right\}&=&\\
		\inf _{\lambda \geq 0, \tilde{\eta}\in\RR}\left\{(k-1)^{k_{*}} k^{-1} \mathbb{E}_{\mu}(\tilde{H}(x,\theta)-\tilde{\eta})_{+}^{k_{*}} \lambda^{1-k_{*}}+\left(\rho+\frac{1}{k(k-1)}\right) \lambda+\tilde{\eta}\right\}&&
	\end{eqnarray*}
	where in the last line followed by setting $\tilde{\eta}:=\eta-\frac{\lambda}{k-1}$. Taking derivatives with respect to $\lambda$ to infimize the preceding expression, we have (noting that $\left.\left(k_{*}-1\right) / k_{*}=1 / k\right)$
	\begin{equation*}
		\lambda=(k-1)(k(k-1) \rho+1)^{-\frac{1}{k_{*}}}\left(\mathbb{E}_{\mu}(\tilde{H}(x,\theta)-\tilde{\eta})_{+}^{k_{*}}\right)^{\frac{1}{k_{*}}}
	\end{equation*}
	
	By substituting into the preceding expression, we find that the infimum is 
	\begin{equation*}
		\inf _{\tilde{\eta}}(k(k-1) \rho+1)^{\frac{1}{k}}\left(\mathbb{E}_{\mu}(\tilde{H}(x,\theta)-\tilde{\eta})_{+}^{k_{*}}\right)^{1 / k_{*}}+\tilde{\eta}.
	\end{equation*}
	
	Defining $m_k(\rho)=(1+k(k-1) \rho)^{\frac{1}{k}}$ the conclusion follows at once.\eproof
	
	\subsection{Proof of Proposition \ref{Robust_Consideration_Sets}} The proof of this result is a simple  adaptation of \cite[Prop. 1]{Caplinetal2019} to the case of program (\ref{Max_Lag_phi_DRO_phi}). Accordingly, the optimization problem has the associated  Lagrangian 
	{\footnotesize
		$$\mathcal{L}(x,\lambda,\eta, \varphi,\kappa)=-\eta-\rho \lambda-\lambda \sum_{\theta\in \Theta} \mu(\theta) \phi^{*}\left(\frac{-H(x, \theta)-\eta}{\lambda}\right)+\varphi\left(1-\sum_{i\in A}x_i\right)+\sum_{i\in A}\kappa_ix_i+\vartheta\lambda,$$}where $\varphi$   is the Lagrangian multiplier on the constraint that the unconditional probabilities $x=(x_1,\ldots,x_n)$ must sum to 1, and $\kappa_{i}$ is the multiplier on the non-negativity constraint for $x_i$ (we use $\kappa_i$ rather than $\xi_i$ here to avoid clashing with the attention-cost parameter $\xi$). Similarly, $\vartheta$ is the Lagrangian multiplier on the non-negativity constraint for $\lambda$ (we write $\vartheta$ to avoid a clash with the penalty weight $\psi$ of \S\ref{s36}).
	\smallskip
	
	Defining $\nu(\theta)\triangleq\mu(\theta)\times\phi^{\ast\prime}\left({-H(x,\theta)-\eta\over \lambda}\right)\geq 0$ for all $\theta\in \Theta$ (non-negative because $\phi^*$ is non-decreasing, as $\operatorname{dom}\phi\subseteq[0,\infty)$), the associated  first order conditions with respect to $\eta,\lambda,$ and $x$ are:
	\begin{eqnarray}
		1-\sum_{\theta\in \Theta}\nu(\theta)&=&0\label{FOC1},\\
		-\rho-\sum_{\theta\in \Theta}\mu(\theta)\phi^{\ast}
		\left({-H(x,\theta)-\eta\over\lambda}\right)&&\nonumber\\
		-{1\over\lambda}\sum_{\theta\in \Theta}\nu(\theta)
		\left(H(x,\theta)+\eta\right)+\vartheta&=&0\label{FOC2},\\
		\xi\sum_{\theta\in \Theta}\nu(\theta)
		\left({e^{u(i,\theta)/\xi}\over\sum_{j\in A}x_je^{u(j,\theta)/\xi}}\right)-\varphi+\kappa_i	&=&0\quad\mbox{for all $i\in A,$}\label{FOC3}
	\end{eqnarray}
	using $\partial H(x,\theta)/\partial x_i=\xi\, e^{u(i,\theta)/\xi}/\sum_{j\in A}x_je^{u(j,\theta)/\xi}$. Since $\nu(\theta)\geq0$ for all $\theta\in \Theta$, (\ref{FOC1}) implies that $\nu$ is a well-defined probability measure on $\Theta$, possibly without full support.

	The complementary slackness conditions on the choice probabilities are $\kappa_{i} x_i=0$, $\kappa_i \geq 0$. If $x_i>0$, then $\xi\sum_{\theta}\nu(\theta)e^{u(i,\theta)/\xi}/\sum_{j}x_je^{u(j,\theta)/\xi}=\varphi$; multiplying by $x_i$, summing over $i$, and using $\sum_{\theta}\nu(\theta)\sum_ix_ie^{u(i,\theta)/\xi}/\sum_jx_je^{u(j,\theta)/\xi}=\sum_\theta\nu(\theta)=1$ gives $\varphi=\xi$, so $x_i>0$ implies, after dividing by $\xi$,
	$$\sum_{\theta\in \Theta}\nu(\theta)
	\left({e^{u(i,\theta)/\xi}\over\sum_{j\in A}x_je^{u(j,\theta)/\xi}}\right)=1,$$
	while $x_i=0$ (with $\kappa_i\geq 0$) gives, after the same division by $\xi$, the corresponding inequality
	$$\sum_{\theta\in \Theta}\nu(\theta)
	\left({e^{u(i,\theta)/\xi}\over\sum_{j\in A}x_je^{u(j,\theta)/\xi}}\right)=1-\kappa_i/\xi\leq 1,$$
	which are necessary and sufficient conditions for an optimal $x\in \Delta_n$.

	In (\ref{FOC2}), the complementary slackness condition on $\lambda$ is $\vartheta\lambda=0$, $\lambda\geq 0$. For $\lambda>0$, (\ref{FOC2}) can be written as
	$$-\rho-\sum_{\theta\in \Theta}\mu(\theta)\phi^{\ast}
	\left({-H(x,\theta)-\eta\over\lambda}\right)=
	{1\over\lambda}\sum_{\theta\in \Theta}\nu(\theta)(H(x,\theta)+\eta),$$
	equivalently
	$$-\eta-\rho\lambda-\lambda\sum_{\theta\in \Theta}\mu(\theta)\phi^{\ast}\left({-H(x,\theta)-\eta\over\lambda} \right)=\sum_{\theta\in \Theta}\nu(\theta)H(x,\theta).$$
	\eproof

	\subsection{Proof of Corollary \ref{Corollary_DRCS}} This is Proposition \ref{Robust_Consideration_Sets} specialized to $\phi=\phi_{kl}$. By Example \ref{Kullback_Leibler}, $\phi_{kl}^{*\prime}(s)=e^s$, so the worst-case prior of Proposition \ref{Robust_Consideration_Sets} is
	\begin{align*}
	\nu(\theta)&=\mu(\theta)\phi_{kl}^{*\prime}\!\left(\frac{-H(x^*,\theta)-\eta^*}{\lambda^*}\right)=\mu(\theta)\,e^{-(H(x^*,\theta)+\eta^*)/\lambda^*}\\
	&=\mu(\theta)\,e^{-\eta^*/\lambda^*}e^{-H(x^*,\theta)/\lambda^*},
	\end{align*}
	and, exactly as in the footnote to Example \ref{Kullback_Leibler}, summing over $\theta$ and imposing $\sum_\theta\nu(\theta)=1$ gives $e^{\eta^*/\lambda^*}=\EE_\mu[e^{-H(x^*,\theta)/\lambda^*}]$, so that
	$$\nu(\theta)=\frac{\mu(\theta)\,e^{-H(x^*,\theta)/\lambda^*}}{\EE_\mu\!\left[e^{-H(x^*,\theta)/\lambda^*}\right]}.$$
	Substituting this expression for $\nu(\theta)$ into condition (\ref{Consideration_Set_Condition}) of Proposition \ref{Robust_Consideration_Sets} yields exactly (\ref{Consideration_Set_Condition2}).\eproof

	\subsection{Proof of Corollary \ref{Corollary_AVaR_DRCS}} By (\ref{concentration_ex1})--(\ref{R_VaR}), $x^*$ solves the joint program
	$$\max_{x\in\Delta_n,\,\eta\in\RR}\left\{-\eta-\frac1\beta\sum_{\theta\in\Theta}\mu(\theta)(\tilde H(x,\theta)-\eta)_+\right\},$$
	which is well-defined since $-\mathcal{R}(x)=\sup_\eta\{-\eta-\beta^{-1}\EE_\mu[(\tilde H(x,\theta)-\eta)_+]\}$. This program is jointly concave in $(x,\eta)$: $\tilde H(x,\theta)=-H(x,\theta)$ is convex in $x$ (Theorem \ref{Thm_DRO-RI-phi}'s proof establishes concavity of $H$), affine in $\eta$, and $(\cdot)_+$ is convex and non-decreasing, so $(\tilde H(x,\theta)-\eta)_+$ is jointly convex in $(x,\eta)$ as the composition of a non-decreasing convex function with a jointly convex one. The associated Lagrangian is
	$$\mathcal{L}(x,\eta,\varphi,\kappa)=-\eta-\frac1\beta\sum_{\theta\in\Theta}\mu(\theta)(\tilde H(x,\theta)-\eta)_++\varphi\Big(1-\sum_{i\in A}x_i\Big)+\sum_{i\in A}\kappa_ix_i,$$
	with $\varphi,\kappa_i$ as in the proof of Proposition \ref{Robust_Consideration_Sets}.

	Consider first the generic case in which no state ties at the threshold, i.e. $\tilde H(x^*,\theta)\neq\eta^*$ for all $\theta\in\Theta$. Then $(\cdot)_+$ is differentiable at every $\tilde H(x^*,\theta)-\eta^*$, with derivative $1\{\tilde H(x^*,\theta)>\eta^*\}$, and the first-order condition in $\eta$ reads
	$$-1+\frac1\beta\sum_{\theta\in\Theta}\mu(\theta)\,1\{\tilde H(x^*,\theta)>\eta^*\}=0,$$
	i.e. $\sum_{\theta\in\Theta}\nu(\theta)=1$ for $\nu$ as in (\ref{AVaR_nu}) (there being no state to split, so the $t_\theta$ play no role), so $\nu$ is a well-defined probability measure on $\Theta$. If instead $\Theta_0\triangleq\{\theta\in\Theta:\tilde H(x^*,\theta)=\eta^*\}$ is non-empty, $(\cdot)_+$ is only subdifferentiable at $0$, with subdifferential $[0,1]$. Because the objective is separable across $\theta$, its subdifferential with respect to $\eta$ at $\eta^*$ is the sum, over $\theta\in\Theta_0$, of the individual subdifferentials $\mu(\theta)[0,1]$: there is no requirement that these state-specific subgradients coincide. Writing $t_\theta\in[0,1]$ for the subgradient selected at state $\theta\in\Theta_0$, the first-order condition in $\eta$ becomes
	$$-1+\frac1\beta\left[\sum_{\theta:\tilde H(x^*,\theta)>\eta^*}\mu(\theta)+\sum_{\theta\in\Theta_0}\mu(\theta)t_\theta\right]=0,$$
	which is exactly $\sum_{\theta\in\Theta}\nu(\theta)=1$ for $\nu$ as in (\ref{AVaR_nu}). This single scalar equation pins down only the $\mu$-weighted sum $\sum_{\theta\in\Theta_0}\mu(\theta)t_\theta$; the individual $\{t_\theta\}_{\theta\in\Theta_0}$ are determined jointly with this normalization by the first-order conditions in $x$ derived below, and need not be unique. This is the discrete-distribution analogue of the standard treatment of ties at the VaR quantile in the Rockafellar--Uryasev representation of AV$@$R (\citealp{RockafellarUryasev}), generalized to allow state-specific mass splitting across a multi-point tie.

	In either case, differentiating $\mathcal{L}$ with respect to $x_i$ gives, by the same chain-rule computation as in the proof of Proposition \ref{Robust_Consideration_Sets} (using $\partial H(x,\theta)/\partial x_i=\xi\, e^{u(i,\theta)/\xi}/\sum_{j\in A}x_je^{u(j,\theta)/\xi}$ and $\partial\tilde H/\partial x_i=-\partial H/\partial x_i$),
	$$\xi\sum_{\theta\in\Theta}\nu(\theta)\left(\frac{e^{u(i,\theta)/\xi}}{\sum_{j\in A}x_j^*e^{u(j,\theta)/\xi}}\right)-\varphi+\kappa_i=0,\qquad i\in A,$$
	with complementary slackness $\kappa_ix_i^*=0$, $\kappa_i\geq0$. Multiplying by $x_i^*$, summing over $i\in A$, and using $\sum_i x_i^*=1$ gives $\varphi=\xi$ exactly as in the proof of Proposition \ref{Robust_Consideration_Sets}; dividing through by $\xi$ and using $\kappa_i\geq0$ yields the stated necessary and sufficient conditions (\ref{Consideration_Set_Condition_AVaR}).\eproof

	\subsection{Proof of Proposition \ref{Prop_AVaR_uniform}} Fix $x\in\Delta_n$ and write $Z(\theta)\triangleq\tilde H(x,\theta)$. By the risk-measure representation in \S\ref{s3}, $\mathcal{R}(x)=\max\{\EE_\nu[Z]:\nu\in\mathcal{B}_\phi(\mu)\}$, the maximum of a linear function over the set (\ref{Uncertainty_CVaR}), $\{\nu\in\Delta(\Theta):\nu\le\mu/\beta\}$. This set grows as $\beta$ decreases, so $\mathcal{R}(x)$ is non-increasing in $\beta$, and $\mathcal{R}(x)\le\max_\theta Z(\theta)$ for every $\beta$ because each feasible $\nu$ is a probability measure. Let $\theta^\dagger\in\arg\max_\theta Z(\theta)$. The point mass $\delta_{\theta^\dagger}$ belongs to (\ref{Uncertainty_CVaR}) if and only if $1\le\mu(\theta^\dagger)/\beta$, which holds whenever $\beta\le\underline\mu\triangleq\min_{\theta\in\Theta}\mu(\theta)>0$. Hence $\mathcal{R}(x)=\max_\theta Z(\theta)=-\min_\theta H(x,\theta)$ for every $x\in\Delta_n$ and every $\beta\le\underline\mu$. This proves the monotone convergence, shows that it is attained at the finite level $\underline\mu$ uniformly in $x$, and gives $\max_{x\in\Delta_n}\{-\mathcal{R}(x)\}=\max_{x\in\Delta_n}\min_{\theta\in\Theta}H(x,\theta)$ for all $\beta\le\underline\mu$.

	In the insurance take-up example, $H(x,\theta_j)=\xi\log[\pi(1+\delta x_j)]$ depends on $x$ only through $x_j$, via the same strictly increasing function for every $j$, so $\min_jH(x,\theta_j)$ is a strictly increasing function of $\min_jx_j$. Since $\min_jx_j\le1/N$ for every $x\in\Delta_n$, with equality if and only if $x_j=1/N$ for all $j$, the uniform vector $x=(1/N,\ldots,1/N)$ is the unique maximizer of $\min_jH(x,\theta_j)$ over $\Delta_n$, independently of $\mu$.\eproof

	\subsection{Proof of Lemma \ref{Lemma_Sbeta}} Fix $K\le N$ and $1\le\alpha\le\alpha'$. Define weights $w_i=1/(K+\delta)$ for $i=1,\ldots,K$ and $w_0=\delta/(K+\delta)$, which sum to $1$, and let $Y$ be the random variable equal to $\mu(\theta_i)$ with probability $w_i$ ($i=1,\ldots,K$) and equal to $0$ with probability $w_0$. Then $T_\alpha(K)=\left(\sum_{i=1}^Kw_i\mu(\theta_i)^{1/\alpha}\right)^\alpha=\left(\EE_w[Y^{p}]\right)^{1/p}$ for $p=1/\alpha$, i.e. $T_\alpha(K)$ is the $L^p(w)$-norm of $Y$. For $0<p\le p'$, Jensen's inequality applied to the convex function $t\mapsto t^{p'/p}$ ($p'/p\ge1$) gives $\left(\EE_w[Y^p]\right)^{p'/p}\le\EE_w[(Y^p)^{p'/p}]=\EE_w[Y^{p'}]$, hence $\left(\EE_w[Y^p]\right)^{1/p}\le\left(\EE_w[Y^{p'}]\right)^{1/p'}$: the $L^p(w)$-norm of $Y$ is non-decreasing in $p$. Since $\alpha\le\alpha'$ implies $p=1/\alpha\ge1/\alpha'=p'$, this gives $T_{\alpha'}(K)=\left(\EE_w[Y^{p'}]\right)^{1/p'}\le\left(\EE_w[Y^{p}]\right)^{1/p}=T_\alpha(K)$.\eproof
	
	\subsection{Proof of Proposition \ref{Consideration_Sets}} Throughout, $\lambda$ denotes the optimal multiplier $\lambda^\star>0$ of Proposition \ref{Robust_Consideration_Sets}. In the constrained model it depends on $\rho$ (and on $x^\star$), so the characterization below holds at the value $\alpha=1+\xi/\lambda^\star$ attained at the optimum; the inclusion $B\subseteq A(x^\star)$ holds whatever that value is, since it uses only $\alpha\geq1$. Substituting $H(x,\theta_i)=\xi\log[\pi(1+\delta x_i)]$ and the Kullback-Leibler worst-case prior $\nu(\theta_i)=\mu(\theta_i)e^{-H(x,\theta_i)/\lambda}/\sum_j\mu(\theta_j)e^{-H(x,\theta_j)/\lambda}=\mu(\theta_i)[\pi(1+\delta x_i)]^{-1/\bar\lambda}/\sum_j\mu(\theta_j)[\pi(1+\delta x_j)]^{-1/\bar\lambda}$ (using $\bar\lambda\triangleq\lambda/\xi$) into the necessary and sufficient conditions of Proposition \ref{Robust_Consideration_Sets} (with $\alpha\triangleq1/\bar\lambda+1$) shows that, for any $i,i'\in A(x^*)$,
	$$
	\frac{\mu(\theta_i)}{(1+\delta x^*_i)^{\alpha}}=\frac{\mu(\theta_{i'})}{(1+\delta x^*_{i'})^{\alpha}}\triangleq c,\qquad i,i'\in A(x^*),
	$$
	while for $k\notin A(x^*)$ the corresponding (weak) version gives $\mu(\theta_k)\le c$.\footnote{Substituting $\nu(\theta_j)=\mu(\theta_j)e^{-H(x,\theta_j)/\lambda}/Z$, with $Z\triangleq\sum_l\mu(\theta_l)e^{-H(x,\theta_l)/\lambda}$, into the condition of Proposition \ref{Robust_Consideration_Sets} and using the payoff transformation (\ref{Payoffs}), both the equality (for $i\in A(x^*)$) and the inequality (for $k\notin A(x^*)$) share the common multiplicative factor $\pi^{1-\alpha}/Z$ and the common additive terms $\sum_{j\in A(x^*)}\mu(\theta_j)(1+\delta x_j^*)^{-\alpha}+\sum_{l\notin A(x^*)}\mu(\theta_l)$; subtracting the excluded-action condition from the included-action condition cancels $Z$ and $\pi$ entirely and isolates the stated comparison $\mu(\theta_k)\lessgtr\mu(\theta_i)(1+\delta x_i^*)^{-\alpha}$.} Writing $K\triangleq|A(x^*)|$, $c$ is pinned down by $\sum_{i\in A(x^*)}x_i^*=1$: since $1+\delta x_i^*=(\mu(\theta_i)/c)^{1/\alpha}$ for $i\in A(x^*)$, summing over $i\in A(x^*)$ gives $K+\delta=c^{-1/\alpha}\sum_{i\in A(x^*)}\mu(\theta_i)^{1/\alpha}$, i.e. $c=\left(\sum_{i\in A(x^*)}\mu(\theta_i)^{1/\alpha}/(K+\delta)\right)^{\alpha}$. The equal-ratio condition also delivers the equality in (\ref{Consideration_Set_Condition}) itself, not only the comparisons between actions: writing $w_j\triangleq1+\delta x_j^*$ (so that $w_j=1$ for $j\notin A(x^*)$), the included-action condition reads $\sum_j\mu(\theta_j)w_j^{-\alpha}+\delta c=\sum_j\mu(\theta_j)w_j^{1-\alpha}$, and this holds automatically because $\sum_j\mu(\theta_j)w_j^{-\alpha}(w_j-1)=\sum_{j\in A(x^*)}\mu(\theta_j)w_j^{-\alpha}\,\delta x_j^*=\delta c\sum_{j\in A(x^*)}x_j^*=\delta c$. Hence the policy in (\ref{Optimal_policy}) satisfies the necessary and sufficient conditions of Proposition \ref{Robust_Consideration_Sets}. Since $i\in A(x^*)$ requires $\mu(\theta_i)=c(1+\delta x_i^*)^\alpha>c$ while $k\notin A(x^*)$ requires $\mu(\theta_k)\le c$, $c$ is exactly the threshold separating included from excluded actions: $x_K^*>0$ requires
	$$\mu(\theta_K) > \left(\frac{\sum_{a_i\in A(x^*)} (\mu(\theta_i))^{1/\alpha}}{\delta+K}\right)^{\alpha}=T_\alpha(K),$$
	and $j\notin A(x^*)$ requires $\mu(\theta_j)\leq T_\alpha(K)$,\footnote{We recall that without loss of generality we have assumed that $\mu(\theta_1)\geq \mu(\theta_2)\geq\ldots\geq \mu(\theta_N)>0$.} establishing the threshold characterization (\ref{Optimal_policy}) and the definition of $T_\alpha(K)$ used in the main text; the only difference from \cite{Caplinetal2019} is that their derivation has $\alpha=1$. Existence and uniqueness of the cutoff $K$ satisfying (\ref{Inequalities_Consideration}) likewise follow from \cite[Thm.~1]{Caplinetal2019}'s single-crossing argument, applied with $\tilde\mu_i\triangleq\mu(\theta_i)^{1/\alpha}$ in place of $\mu(\theta_i)$: taking $1/\alpha$-th powers, $\mu(\theta_k)>T_\alpha(k)$ if and only if $\tilde\mu_k>\sum_{i\le k}\tilde\mu_i/(k+\delta)$, which is their condition with weights $\tilde\mu_i$. This condition is homogeneous of degree one in $(\tilde\mu_i)$, so the normalization $\sum_i\tilde\mu_i=1$ plays no role, and their argument uses only that the $\tilde\mu_i$ are positive and non-increasing in $i$ (which holds since $t\mapsto t^{1/\alpha}$ is increasing); it therefore goes through unchanged for any fixed $\alpha\ge1$, and shows that $\{K:\mu(\theta_K)>T_\alpha(K)\}$ is an initial segment $\{1,\ldots,K\}$ of $\{1,\ldots,N\}$.
	
	We now prove the set-inclusion claim $B\subseteq A(x^*)$. Write $K_1$ for the cutoff at $\alpha=1$, so $B=\{\theta_1,\ldots,\theta_{K_1}\}$, and $K$ for the cutoff at the given $\alpha\ge1$, so $A(x^*)=\{\theta_1,\ldots,\theta_K\}$. Suppose, for contradiction, that $K<K_1$. Since $K+1\le K_1$ and $\{1,\ldots,K_1\}$ is the initial segment on which the $\alpha=1$ condition holds, $\mu(\theta_{K+1})>T_1(K+1)$. But $K$ being the $\alpha$-cutoff and $K+1>K$ means $\mu(\theta_{K+1})\le T_\alpha(K+1)$. Combining the two displays, $T_\alpha(K+1)\ge\mu(\theta_{K+1})>T_1(K+1)$, contradicting Lemma \ref{Lemma_Sbeta} (which requires $T_\alpha(K+1)\le T_1(K+1)$ since $\alpha\ge1$). Hence $K\ge K_1$, so $A(x^*)=\{\theta_1,\ldots,\theta_K\}\supseteq\{\theta_1,\ldots,\theta_{K_1}\}=B$.\eproof

\end{document}